\documentclass[pra,twocolumn,showpacs,superscriptaddress]{revtex4}
\usepackage{graphicx,graphics,psfrag,amsmath,calc,amssymb}
\usepackage{color}

\begin{document}
\title{Confinement-induced Resonances in Quasi-one-dimensional Traps with Transverse Anisotropy}
\author{Wei Zhang}
\author{Peng Zhang}
\affiliation{Department of Physics, Renmin University of China, Beijing, 100872, China}

\date{\today}

\begin{abstract}

We study atom-atom scattering in quasi-one-dimensional geometries with transverse anisotropy.
By assuming an $s$-wave pseudo-potential of contact interaction, we show that the system would exhibit a
single confinement-induced resonance, where the scattering process degenerates to a total reflection
as a one-dimensional gas of impenetrable bosons. For a general form of interaction, we present a
formal calculation based on the two-channel model and conclude the existence of only one confinement-induced
resonance. Our findings are inconsistent with a recent experiment by Haller {\it et al.}
[Phys. Rev. Lett. 104, 153203 (2010)], where a splitting of confinement-induced resonances
has been observed in an anisotropic quasi-one-dimensional quantum gas of Cs atoms.

\end{abstract}

\maketitle

%
%

\section{Introduction}
\label{sec:introduction}

The interest on low-dimensional systems has been recently intensified
by the experimental development of manipulating ultracold quantum gases in
optical lattices~\cite{stoferle-04, Kinoshita-04, Paredes-04, Moritz-05, Stock-05,
Gunter-05, Chin-06, Hadzibabic-06} and on atom chips~\cite{Zimmermann-07}.
With the aid of tuning an external magnetic field through a Feshbach resonance,
these techniques provide a fascinating possibility of investigating distinct physical
properties in quasi-low-dimensionality with a controllable particle-particle
interaction. Up to now, the Berezinskii-Kosterlitz-Thouless (BKT) transition
has been observed in a series of quasi-two-dimensional (quasi-2D) pancake shaped
Bose gases~\cite{Hadzibabic-06}, and the strongly interacting Tonks-Girardeau (TG)
Bose gas has been achieved in a setup of 2D optical
lattices~\cite{Kinoshita-04, Paredes-04, Syassen-08, Haller-09}.
Of particular interest is the quasi-one-dimensional (quasi-1D) regime,
where only the ground state of transverse motion is significantly populated.
This regime is important partly due to its theoretical simplicity. In fact,
the homogeneous system of 1D Bose gas with $\delta$-contact interaction
is known as one of the few fully integrable quantum systems~\cite{Lieb-63}.
In a finite system with infinitely strong $\delta$ interactions, a $N$-body Bose
system has been proved to correspond via a one-to-one mapping with the highly
correlated states of the corresponding noninteracting Fermi
gas~\cite{Girardeau-63, Tonks-36, Girardeau-01, Dunjko-01, Petrov-00a, Andersen-02}.

The quasi-1D geometry can be realized by arranging a 2D optical lattice in
the transverse $(x$-$y)$ plane, such that the kinetic and the interaction energy of the particles
are insufficient to transfer particles to transversally excited energy levels.
However, even in this ultimate limit, one should bear in mind that the transverse degrees
of freedom are of great importance and could manifest themselves in many aspects of
the system properties. For example, when the particle-particle interaction can support
a bound state, the binding energy will serve as an additional energy scale, which can
be comparable or even exceed the transverse confinement. In this situation, the transversally
excited states will be inevitably populated, and have to be taken into account for a proper
description of the system~\cite{Kestner-06, Zhang-08}. On the other hand, when the
interaction does not support a bound state, the level structure of transversally
energy states will provide extra channels for the 1D scattering process.
The presence of these channels gives a scheme to tune through the 1D scattering
resonance by either changing the three-dimensional (3D) $s$-wave scattering length,
or varying the transverse confinement, which is known as confinement-induced
resonance (CIR) as proposed by Olshanii~\cite{Olshanii-98}.
Around the resonance point of CIR, the effective 1D coupling constant can be varied from
$-\infty$ to $+\infty$, and the 1D scattering process degenerates to a total reflection,
thereby creating a gas of impenetrable bosons.

The underlying physics of CIR is in the same spirit of a Feshbach resonance
between open and closed scattering channels~\cite{Bergeman-03}.
In the presence of transverse confinement, the scattering of atoms in the
transversally ground state assumes the open channel, while the transversally
excited states {\it as a whole} can support a bound state and serve as the closed
channel. According to the Feshbach scheme, the resonance takes place when the
bound state of the closed channel degenerates with the continuum threshold of
the open channel, which predicts the location where CIR occurs.
By assuming a 3D Fermi-Huang pesudo-potential for the particle interaction,
it is shown that the CIR takes place in a cylindrically symmetric quasi-1D harmonic trap
at a specific ratio of $a_s/a_\perp = - 1/\zeta(1/2)$, where $a_s$ is the 3D $s$-wave scattering length,
$a_\perp = \sqrt{\hbar / \mu \omega_\perp}$ is the length scale of transverse harmonic
oscillator with trapping frequency $\omega_\perp = \omega_x = \omega_y$,
$\mu$ is the reduced mass of the two colliding particles, and $\zeta(x)$ is the
Riemann Zeta function~\cite{Olshanii-98}. Further theoretical investigation extends the
discussion of CIR to pancake shaped quasi-2D geometry~\cite{Petrov-00b},
to general forms of interactions~\cite{Kim-05, Naidon-07, Saeidian-08},
to the three-body~\cite{Mora-04} and the four-body~\cite{Mora-05} scattering
processes, as well as to a pure $p$-wave scattering of fermions~\cite{Granger-04}.

Experimental evidence of CIRs has been recently reported for
bosonic~\cite{Kinoshita-04, Paredes-04,Haller-09, Haller-10}
and fermionic~\cite{Gunter-05} systems. In particular, an experiment by
Haller {\it et al.} performs a systematical investigation on CIRs in a ultracold quantum gas of Cs atoms
confined in a quasi-1D geometry with transverse anisotropy~\cite{Haller-10}. By varying
the 3D $s$-wave scattering length around a magnetic Feshbach resonance, they determined
the positions of CIR from atom loss and heating rate measurements. Of particular interest
is the CIR was found to split into two resonances upon introducing anisotropy to the transverse
confinement $\omega_x \neq \omega_y$. One of the resonances, CIR1,
shows a pronounced shift to higher values of $a_s/a_y$ by increasing anisotropy
$\eta \equiv \omega_x/\omega_y$. The second resonance, CIR2, shifts instead
to lower values of $a_s/a_y$. Here, $a_y = \sqrt{\hbar/(\mu \omega_y)}$
is the harmonic oscillator length in the $y$-direction.
\begin{figure}
\begin{center}
\psfrag{as}{$a_s/a_s(\eta=1)$}
\psfrag{eta}{$\eta = \omega_x /\omega_y$}
\includegraphics[width=8.5cm]{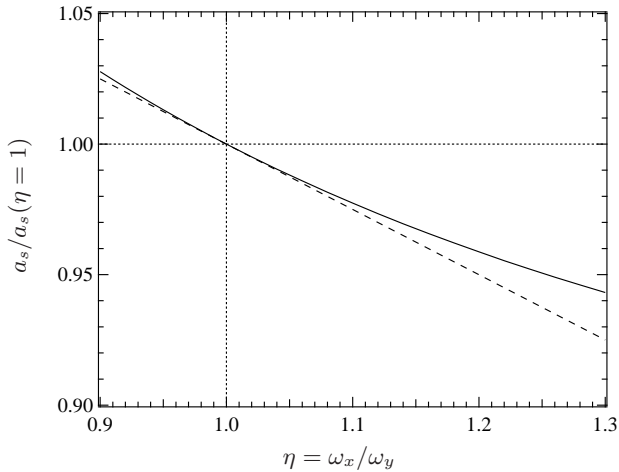}
\caption{Locations of CIRs as a function of the transverse anisotropy
$\eta \equiv \omega_x / \omega_y$. By assuming a 3D Fermi-Huang pseudo-potential,
we find a single CIR for all values of transverse anisotropy (solid).
For small anisotropy of $\eta \approx 1$, the resonance position shifts
downwards with increasing $\eta$ at a slope of $-1/4$ (dashed).}
\label{fig:CIRzoom}
\end{center}
\end{figure}

In this manuscript, we present a theoretical investigation on the CIR in a quasi-1D harmonic trap
with transverse anisotropy. By using a 3D Fermi-Huang pseudo-potential, we show
that only one {\it single} CIR would be present for all values of anisotropy ratio $\eta$.
For small transverse anisotropy, the resonance position of $a_s /a_y$ shifts downwards
with increasing $\eta$. This is in the same trend as the experimental observation of
CIR2~\cite{Haller-10}, but is systematically above the measured values.
The appearance of a single CIR can be well understood by noticing that the $s$-wave
contact interaction can only support {\it one} bound state in the closed channel of excited
transverse modes, and the resonance can only occur when this bound state energy coincide
with the threshold in the open channel of ground transverse mode.
In order to further validate the observation to a general form of $s$-wave interaction,
we imply a formal scattering theory of a two-channel model, and draw the same
qualitative conclusion of a {\it single} CIR. Upon these findings, we conclude that the
splitting of peaks of atom loss and heating rate observed in the experiment cannot
be simply understood by the theory of CIR.

The remaining of this manuscript is organized as follows. In Sec.~\ref{sec:scattering}, we discuss
the $s$-wave scattering process of two atoms in a quasi-1D geometry with transverse anisotropy.
By assuming a $\delta$-contact pseudo-potential, we calculate the position of CIR for
a wide range of transverse anisotropy. This result is then confirmed by another calculation of
the closed channel bound state energy, as discussed in Sec.~\ref{sec:bound}. Considering the fact
that the CIR takes place at the exact position where the bound state of the closed channel
degenerates to the threshold of the open channel, we can obtain resonance condition from a
different aspect. In Sec.~\ref{sec:formal theory}, we extend the discussion to a general form
of interaction, and present a formal theory for two-atom scattering within a two-channel model.
Finally, we summarize our conclusions in Sec.~\ref{sec:conclusion}. We put some
detail calculations for Sec.~\ref{sec:formal theory} in Appendix \ref{sec-appendix-D}, 
while some necessary calculations for Appendix \ref{sec-appendix-D} are
posted in Appendix \ref{sec-appendix-A}-\ref{sec-appendix-C}.


\section{Hamiltonian and two-body scattering problem}
\label{sec:scattering}

We consider the problem of two interacting atoms with mass $m$ colliding in
a 2D harmonic confinement in the transverse $x$-$y$ plane
\begin{equation}
\label{1-trap}
U({\bf r}) = \frac{m}{2} \left( \omega_x^2 x^2 + \omega_y^2 y^2\right).
\end{equation}
In general, the transverse trapping frequencies $\omega_{x,y}$ along the $x$ and $y$ directions
are different, with transverse anisotropy ratio $\eta = \omega_x / \omega_y$.
The atomic motion along the $z$-axis is assumed to be free. Then the Hamiltonian of the system 
is given by
\begin{equation}
\label{1-Hamiltonian}
{\hat H} = -\frac{\hbar^2}{2 m} \nabla_{{\bf r}_1}^2  - \frac{\hbar^2}{2 m} \nabla_{{\bf r}_2}^2
+ U({\bf r}_1) + U({\bf r}_2) + V({\bf r}_1 - {\bf r}_2),
\end{equation}
where ${\bf r}_1$ and ${\bf r}_2$ denote the positions of the two atoms, respectively.
$V({\bf r}_1 - {\bf r}_2)$ is the interaction potential between the two atoms.
For sufficiently low energies, the 3D atom-atom scattering is dominated by $s$-wave 
scattering. Therefore, in the present and the next
sections, we model the interaction potential $V$ by a Fermi-Huang-Yang pseudo-potential \cite{pseudopotential}
in the form of regularized $\delta$-function
\begin{equation}
\label{1-pseudo-potential}
V({\bf r}) = \frac{4 \pi \hbar^2 a_s}{m} \delta({\bf r}) \frac{\partial}{\partial r} r,
\end{equation}
where $a_s$ is the 3D $s$-wave scattering length. The discussion of a general form of interaction
will be left in Sec.~\ref{sec:formal theory}.

One great advantage of harmonic trapping potential is the center-of-mass (COM) and
relative degrees of freedom can be decoupled. Substituting new coordinates
${\bf r} = {\bf r}_1 - {\bf r}_2$ and ${\bf R} = ({\bf r}_1 + {\bf r}_2)/2$,
we separate the full Hamiltonian into the COM part ${\hat H}_{\rm COM}$ and
the relative part ${\hat H}_{\rm rel}$
\begin{eqnarray}
{\hat H}_{\rm COM} &=& -\frac{\hbar^2}{2M }\nabla_{\bf R}^2 + 2 U({\bf R}),
\\
{\hat H}_{\rm rel} &=&  -\frac{\hbar^2}{2 \mu }\nabla_{\bf r}^2 + \frac{U({\bf r})}{2} + V({\bf r}),
\end{eqnarray}
where $M = 2m$ and $\mu = m/2$ are the total and reduced masses, respectively.
The eigenfuctions of the COM Hamiltonian is simply a combination of harmonic oscillators
in the transverse plane and plane waves along the $z$-direction, which can be safely
dropped out from our discussion. The relative Hamiltonian can be expressed in a
dimensionless form, in which all lengths are scaled in units of
$a_y = \sqrt{\hbar / (\mu \omega_y)}$, and all energies are scaled in units of
$\hbar \omega_y$. The eigenfunctions $\Psi({\bf r})$ of the relative motion thus are
governed by the Schr{\"o}dinger equation
\begin{eqnarray}
\label{1-Schrodinger}
&&\left[
-\frac{1}{2} \nabla_{\bf r}^2 + \frac{1}{2}(\eta^2 x^2 + y^2)
+ 2 \pi a_s \delta({\bf r}) \frac{\partial}{\partial r}r
\right]
\Psi({\bf r})
= E \Psi({\bf r}),\nonumber\\
&&
\label{eigen}
\end{eqnarray}
where the eigenenergy $E = (k_z^2 + \eta +1)/2$ is the summation of the
longitudinal kinetic energy $k_z^2$ and the ground-state energy of the
transverse confinement. In this paper we assume 
the incident wave function is in the transverse ground state 
and longitudinal kinetic
energy is smaller than energy spacing between the ground and first excited transverse
mode, i.e., $k_z^2<{\rm min}\{\eta,1\}$. In this case 
the asymptotic scattering-state wave function in the limit $|z| \to \infty$
would be definitely in the transverse ground state.

As a result of the zero-range pseudo potential $V(\bf{r})$ in (\ref{1-pseudo-potential}), the scattering state with 
odd parity along the $z$ direction does not get any phase shift. Then the relevant scattering
amplitude is zero. Thus, we we focus on the scattering state with even parity in the following discussion.

In the region ${\bf r}\neq 0$, the pseudo-potential $V({\bf r})$ 
takes zero value. Therefore the atomic motion with respect to 
different eigenstates of the transverse confinement are decoupled with each other. Then
 the eigen-equation (\ref{eigen}) can be solved analytically.
The solution
with even parity and out-going boundary condition in the $z$ direction takes the form
\begin{eqnarray}
&& \Psi(x,y,z) =
\left(\cos(k_z z) + f_e e^{i k_z |z|} \right) \eta^{1/4} \Phi_{0}(\sqrt{\eta} x) \Phi_{0}(y)
\nonumber \\
&& \hspace{0mm}
+ \sum_{(n_x, n_y)'} B_{n_x n_y} \eta^{1/4} \Phi_{n_x}(\sqrt{\eta} x) \Phi_{n_y}(y)
e^{- \kappa_{n_x n_y} |z|}.\label{solu}
\end{eqnarray}
where $\kappa_{n_x, n_y} = \sqrt{2 n_x \eta + 2 n_y - k_z^2}$,
and the summation runs over all combinations of $(n_x, n_y)$ excepts the
$(n_x =0, n_y =0)$ ground mode.
In the expression above $\eta^{1/4}\Phi_n(\sqrt{\eta}x)$ and $\Phi_n(y)$ are the normalized
 wave functions of the $n$-th excited states
of the confinements in the $x$ and $y$ direction, respectively, 
with $\Phi_n(t)$ the dimensionless wave function of harmonic oscillator
\begin{eqnarray}
\Phi_n (t) = \frac{e^{-t^2/2}}{\pi^{1/4} \sqrt{2^n n!}} H_n(t). \label{1-harmonics}
\end{eqnarray}
Here, $H_n(t)$ is the Hermite polynomial.

The scattering amplitude $f_e$ and the parameters $B_{n_x n_y}$ can be formally extracted 
from the connection condition given by the $s$-wave pseudo potential $V({\bf r})$ 
at the origin ${\bf r}=0$. To this end we substitute Eq. (\ref{solu}) into Eq. (\ref{eigen}),
and do the integration 
\begin{eqnarray}
\lim_{\varepsilon\rightarrow 0}\int_{-\varepsilon}^{+\varepsilon}dz
\iint_{-\infty}^{+\infty}dx dy
\left[\eta^{1/4} \Phi^{\ast}_{n_x}(\sqrt{\eta} x) \Phi^{\ast}_{n_y}(y)\right]
\end{eqnarray}
on both sides of the equation. Then we get
\begin{eqnarray}
f_e &=& \frac{2 \pi a_s}{i k_z} \eta^{1/4} | \Phi_{0}(0) |^2 {\cal S},
\label{1-amplitude}
\\
B_{n_x n_y} &=& -\frac{2 \pi a_s}{\kappa_{n_x n_y}}
\eta^{1/4} \Phi_{n_x}(0)  \Phi_{n_y}(0) {\cal S},
\label{1-coeff}
\end{eqnarray}
where the factor ${\cal S}$ is defined as
\begin{equation}
\label{1-S}
{\cal S} = \left[\frac{\partial}{\partial r} r \Psi({\bf r}) \right]_{r \to 0} =
\left[\frac{\partial}{\partial z} z \Psi(0,0,z) \right]_{z \to 0^+}.
\end{equation}

To derive the finial expression of the scattering amplitude, we
substitute Eqs. (\ref{1-amplitude}) and (\ref{1-coeff}) into Eq. (\ref{solu}),
and express $\Psi(0,0,z)$ in terms of the parameter ${\cal S}$. Substituting
this result into Eq. ({\ref{1-S}}), we get the equation for ${\cal S}$:
\begin{eqnarray}
\mathcal{S}&=&\left( 1+\frac{2\pi a_{s}}{ik_{z}}\eta ^{1/4}\left\vert \Phi
_{0}\left( 0\right) \right\vert ^{2}\mathcal{S}\right)\eta ^{1/4}\left\vert \Phi
_{0}\left( 0\right) \right\vert ^{2}\nonumber\\
 &&-a_{s}\eta
^{1/2}\beta \left( k_{z},\eta \right) \mathcal{S} \label{1-seq}
\end{eqnarray}
where $\beta(k_z,\eta) = [ \partial_z z \Lambda(z, k_z,\eta)]_{z \to 0^+}$
is the regular part of the series
\begin{equation}
\Lambda(z, k_z,\eta) = 2 \pi
\sum_{(n_x, n_y)'}
\frac{|\Phi_{n_x}(0)|^2 |\Phi_{n_y}(0)|^2 }{\kappa_{n_x n_y}} e^{- \kappa_{n_x n_y} z}.
\label{1-Lambda}
\end{equation}
Solving ${\cal S}$ from Eq. (\ref{1-seq}), we finally obtain the quasi-1D scattering amplitude
\begin{equation}
f_e = \frac{-1}{1- i k_z  [ a_s^{-1} \eta^{-1/2} + \beta(k_z,\eta) ]/2}.
\label{1-amplitude2}
\end{equation}
In the low energy limit of $k_z \to 0$, the result of scattering amplitude
$f_e$ (\ref{1-amplitude2}) can be approximated by a scattering amplitude
\begin{equation}
f_e^\delta (k_z) = \frac{-1}{1+ i k_z a_{\rm 1D}}
\end{equation}
for a 1D $\delta$-potential $ V_{\rm 1D} = g_{\rm 1D} \delta(r) $
of coupling strength $g_{\rm 1D} = - a_{\rm 1D}^{-1} $,
where the 1D scattering length takes the form
\begin{equation}
a_{\rm 1D} = - \frac{1}{2} \left[ \frac{1}{\sqrt{\eta} a_s } + \beta(0, \eta) \right].
\end{equation}
The CIR takes place at the position of $a_{\rm 1D}$ approaches zero,
hence indicating a complete reflection of the scattering amplitude. By
introducing back the physical unit, we obtain the CIR condition
\begin{equation}
\label{1-CIR}
\frac{a_s}{a_y} = - \frac{1}{\sqrt{\eta} \beta(0, \eta)}.
\end{equation}
From this result, we can see clearly that only one single CIR would
be present for arbitrary values of transverse anisotropy ratio $\eta$.

We stress that the partial derivative $\partial/\partial z$ and the summation over
transversally excited states $(n_x, n_y)'$ in $\Lambda$ cannot be
interchanged because the series does not converge uniformly, and the
summation must be done first. This is related to the divergent behavior
of the Green's function at small $r$, which is regularized by the operator
$\partial_r (r \cdot)$. Thus, to determine the positon of CIR, we generally need
to numerically evaluate the series $\Lambda(z,0,\eta)$ for a set of $z$,
and extract the coefficient $\beta(0,\eta)$ through the fit
\begin{equation}
\label{1-fit}
\Lambda(z, 0,\eta) = \frac{\alpha(0,\eta)}{z} + \beta(0,\eta) + \gamma(0,\eta) z + \cdots.
\end{equation}
In order to do this, we rewrite the series as
\begin{equation}
\Lambda(z,0,\eta) = \sum_{(s_x, s_y)'}
P_{s_x} P_{s_y}
\frac{e^{-2 \sqrt{s_x \eta + s_y} z}}{\sqrt{s_x \eta + s_y}},
\end{equation}
where $s_x$ and $s_y$ are non-negative integers and the summation
runs over all possible combinations except the ground state
$(s_x = 0, s_y = 0)$. The coefficient is
\begin{equation}
P_{s_i} = \frac{(2 s_i)!}{2^{2 s_i} (s_i !)^2}
= \frac{\Gamma(s_i+1/2)}{\sqrt{\pi} \Gamma(s_i+1)},
\end{equation}
where $\Gamma(x)$ is the Euler Gamma function.

Before discussing the general cases with arbitrary $\eta$, we first focus on
the cylindrically symmetric trap with $\eta =1$, where analytical result
is available. In this case, we define a new summation variable
$s = s_x + s_y$ and get
\begin{equation}
\label{1-Lambda2}
\Lambda(z,0,\eta=1) = \sum_{s=1}^{\infty}
\frac{e^{-2 \sqrt{s} z}}{\sqrt{s}}.
\end{equation}
To reach this expression, we have utilized the relation
\begin{equation}
\sum_{m=0}^{n}
\frac{\Gamma(m+1/2)}{\Gamma(m+1)}
\frac{\Gamma(n-m+1/2)}{\Gamma(n-m+1)}
= \pi
\end{equation}
for $n= 1,2,3,\dots$ being positive integers. The series of Eq. (\ref{1-Lambda2})
has been evaluated~\cite{Olshanii-98}, leading to the well-known result of
$\beta(0,\eta=1) = \zeta(1/2) \approx - 1.4603\dots$, and the CIR condition reads
\begin{equation}
\label{1-CIR4}
\frac{a_s}{a_y} = - \frac{1}{\zeta(1/2)}.
\end{equation}
\begin{figure}
\begin{center}
\psfrag{as}{$a_s/a_y$}
\psfrag{eta}{$\eta = \omega_x /\omega_y$}
\includegraphics[width=8.5cm]{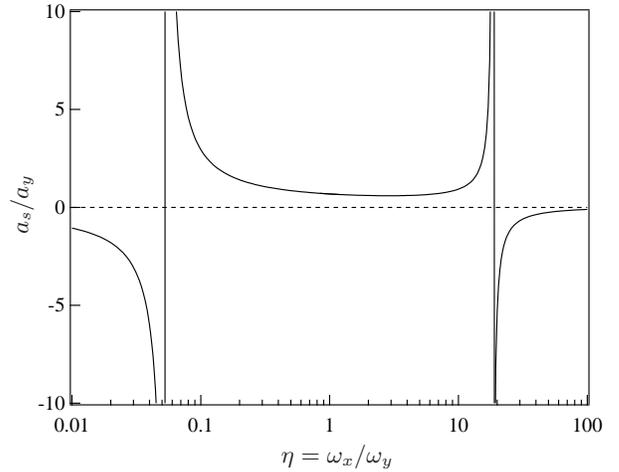}
\caption{Locations of CIR as a function of the transverse anisotropy $\eta = \omega_x / \omega_y$.
By varying $\eta$ from the cylindrically symmetric case $\eta = 1$, the CIR
position shows non-monotonic behavior and crosses over the magnetic Feshbach
resonance to reach the BCS regime with negative scattering length (vertical lines).
}
\label{fig:CIRfull}
\end{center}
\end{figure}

Next, we discuss the general case of $\eta \neq 1$. For a trap with
slight transverse anisotropy $\Delta \eta = \eta -1 \ll 1$, we can expand
the function $\Lambda$ in Eq. (\ref{1-Lambda}) to linear order of $\Delta \eta$.
Notice that the series in uniformly convergent for all finite $\eta$, thus
it is safe to interchange the summation and the partial derivative of $\eta$,
leading to
\begin{equation}
\Lambda(z,0,\eta) \approx  \Lambda(z,0,1) + \Delta \eta \cdot
\frac{\partial \Lambda}{\partial \eta}\bigg\vert_{\eta =1}
+ {\cal O}(\Delta \eta^2)
\end{equation}
with the coefficient
\begin{equation}
\frac{\partial \Lambda}{\partial \eta}\bigg\vert_{\eta =1}
=
- \frac{1}{4} \sum_{s=1}^{\infty}
\left( \frac{1}{\sqrt{s}} + 2 z \right)
e^{-2 \sqrt{s} z}.
\end{equation}
The first term in the bracket corresponds to the same summation
as in Eq. (\ref{1-Lambda2}), and the second term can be evaluated
as
\begin{equation}
\sum_{s=1}^\infty 2z e^{-2 \sqrt{s z^2}} =
\frac{1}{z} + {\cal O}(z), \quad z \to 0^+,
\end{equation}
where we have used the relation
\begin{equation}
\sum_{s=1}^\infty 2 z^2 e^{-2 \sqrt{s z^2}}
\xrightarrow{z \to 0^+}
2 \int_0^\infty dx e^{-2 \sqrt{x}} + {\cal O}(z^2).
\end{equation}
As a consequence, we reach the expression of $\Lambda$ for
$\eta \sim 1$
\begin{equation}
\Lambda(z\to 0^+,0,\eta)
\approx \frac{2 - \Delta \eta }{2 z} + \zeta(1/2)\left(1 - \frac{\Delta \eta}{4}\right) + {\cal O}(z).
\end{equation}
Substituting the expression above into Eq. (\ref{1-CIR}), we get the CIR condition
\begin{equation}
\frac{a_s}{a_y} \approx -\frac{1}{\zeta(1/2)} \left( 1- \frac{\eta-1}{4} \right),
\end{equation}
which indicates that the position of CIR would shift downwards with a slope
of $-1/4$ if we introduce a slight transverse anisotropy by increasing
$\eta = \omega_x/\omega_y$ (see Fig. 1). This result is in the same qualitative trend as
the experimental measurement of CIR2~\cite{Haller-10}, but with
a slope only about half of the experimental observations.

On further increasing or decreasing the transverse anisotropy, the position
of CIR can be determined by numerical evaluation and fit $\Lambda$ to Eq. (\ref{1-fit}).
The results are plotted in Fig. 2 for a wide range of $\eta$. It is shown that the resonance
position of $a_s / a_y$ acquires a non-monotonic behavior of $\eta$, and diverges
at a critical frequency ratio $\eta_c \approx 18.5$
(or equivalently $\eta_c^{-1} \approx 0.054$), where the 3D scattering length crosses through
the magnetic Feshbach resonance to the BCS side of $a_s <0$. This behavior
can be understood by noticing that by continuously imposing the transverse anisotropy,
the quasi-1D geometry will eventually crosses over to a quasi-2D system, where
the $s$-wave CIR occurs at the BCS side of the magnetic Feshbach resonance~\cite{Petrov-00b, Naidon-07}.

\section{Two-channel model and bound state energy}
\label{sec:bound}

In the previous section, we present a study on two-body $s$-wave
scattering process in a transversely anisotropic quasi-1D geometry. We find
a CIR as in a cylindrically symmetric quasi-1D trap. Reminding that
CIR is a Feshbach resonance between the open channel of the
transversally ground state and the closed channel of exited states,
the uniqueness of CIR is rooted from the fact that only one bound state
can be supported by the $s$-wave pseudo-potential. Thus, it is illuminating
to study the bound states of the closed channel and examine the CIR
condition from the binding energy.

In order to extract the detailed information of the eigenstate structure, we
add an additional harmonic trap along the axial $z$-direction with frequency $\omega_z$,
and take the limit of $\omega_z \to 0$ afterwards. This procedure allows us
to take advantage of the simple form of harmonic oscillators. The dimensionless
Hamiltonian of the relative motion of two interacting particle takes the form
\begin{equation}
\label{2-Hamiltonian}
{\hat H}_{\rm rel} = -\frac{1}{2} \nabla_{\bf r}^2 + \frac{1}{2}(\eta^2 x^2 + y^2 + \epsilon^2 z^2)
+ 2 \pi a_s \delta({\bf r}) \frac{\partial}{\partial r}(r \cdot).
\end{equation}
Here, we use $a_y = \sqrt{\hbar / (\mu \omega_y)}$ as the length unit and $\hbar \omega_y$
as the energy limit. By taking the limit of $\epsilon = \omega_z / \omega_y \to 0$, we will reduce
to the original problem with no axial confinement.

Since the CIR is in fact a Feshbach resonance between transversally
ground mode and the manifold of excited modes, we need to investigate
the relevant bound state of the decoupled excited manifold and ask that
when this bound state energy coincides with the continuum threshold of
the ground transverse mode. Thus, we proceed by formally splitting the Hamiltonian
onto "ground" ($g$), "excited" ($e$), and "ground-excited coupling" ($ge$)
parts according to
\begin{eqnarray}
\label{2-twochannel}
{\hat H}_{\rm rel} &=& {\hat H}_{g} + {\hat H}_{e} + {\hat H}_{ge}
\nonumber \\
&=& {\hat P}_{g} {\hat H} {\hat P}_{g} + {\hat P}_{e} {\hat H} {\hat P}_{e}
+ ({\hat P}_{g} {\hat H} {\hat P}_{e} + {\hat P}_{e} {\hat H} {\hat P}_{g}),
\end{eqnarray}
where ${\hat P}_{g} = |00 \rangle_\perp \langle00|$,
${\hat P}_{e} = \sum_{(n_x,n_y)'} |n_x n_y\rangle_\perp \langle n_x n_y|$
are the corresponding projection operators, $|n_x n_y \rangle_\perp$ represents
the eigenstate of the transverse harmonic oscillators with quantum number
$n_x$ and $n_y$, and the summation $(n_x, n_y)'$ runs over all possible
combinations except the ground mode $|00 \rangle_\perp$.

The ground Hamiltonian ${\hat H}_g$ has a 1D coordinate representation,
which corresponds to the motion of a one-dimensional particle in the presence
of a $\delta$ barrier and an axial harmonic trap. In the limit of $\epsilon \to 0$,
the spectrum of ${\hat H}_g$ is continuous for energies above the threshold
energy $E_{g0} = (\eta + 1)/2$, which represents the zero-point energy of
the transverse mode. In the same limit, the spectrum of the excited Hamiltonian
${\hat H}_e$ is also continuous for energies greater than $E_{e0} = 3(\eta +1)/2$.
But as we will see below, ${\hat H}_e$ also supports a bound state with energy
$E_{eb} < E_{e0}$ for all values of 3D scattering length $a_s$.
According to the Feshbach scheme, a resonance would occur when the
bound state energy of ${\hat H}_e$ coincides with the continuum threshold of ${\hat H}_g$,
leading to the resonance condition
\begin{equation}
\label{2-CIR}
E_{eb} = E_{g0} = \frac{\eta +1}{2}.
\end{equation}

The bound state energy $E_{eb}$ of ${\hat H}_e$ can be found by projecting the total
wave function $\Psi$ onto the excited Hilbert space $\Psi_{e} = {\hat P}_e \Psi$,
and decomposing into the harmonic oscillators of the excited Hilbert
space~\cite{Busch-98, Idziaszek-06}
\begin{equation}
\label{2-Psi}
\Psi_e ({\bf r}) = \sum_{(n_x,n_y)'} \sum_{n_z} c_{n_x n_y n_z}
\Phi_{n_x}(\sqrt{\eta} x) \Phi_{n_y} (y) \Phi_{n_z} (\sqrt{\epsilon} z),
\end{equation}
where $\Phi_n(t)$ are the dimensionless harmonic oscillators as given by
Eq. (\ref{1-harmonics}). Substituting the expansion (\ref{2-Psi}) into
the Schr{\"o}dinger equation ${\hat H} \Psi_{e}= E \Psi_{e}$ yields
\begin{eqnarray}
\label{2-eqn}
&& 0 = \sum_{(n_x, n_y)'} \sum_{n_z} c_{\bf n} (E_{\bf n} - E)
{\cal P}_{\bf n} (x,y,z)
\nonumber \\
&& \hspace{0.5cm}
+ 2 \pi a_s \delta({\bf r}) \frac{\partial }{\partial r} r
\sum_{(n_x, n_y)'} \sum_{n_z} c_{\bf n}
{\cal P}_{\bf n} (x,y,z)
\end{eqnarray}
where
\begin{equation}
{\cal P}_{\bf n} (x,y,z) \equiv \Phi_{n_x}(\sqrt{\eta} x)
\Phi_{n_y} (y) \Phi_{n_z} (\sqrt{\epsilon} z)
\end{equation}
Here,  ${\bf n}$ is a shorthand notation for $(n_x, n_y, n_z)$, and
$E_{\bf n} = (n_x+1/2) \eta + (n_y+1/2) + (n_z+1/2) \epsilon$ is the
eigenenergies of 3D harmonic oscillators. To determine the expansion
coefficients $c_{\bf n}$ we project Eq. (\ref{2-eqn}) onto state
$\Phi_{n_x'} \Phi_{n_y'} \Phi_{n_z'}$ with arbitrary $n_x'$, $n_y'$
and $n_z'$, leading to
\begin{equation}
\label{2-coeff}
c_{\bf n} = \frac{2 \pi a_s \sqrt{\eta \epsilon}
{\cal P}_{\bf n} (0,0,0)  {\cal C}}
{E - E_{\bf n}},
\end{equation}
where ${\cal C}$ is a constant fixed by normalization of the wave function.
The value of ${\cal C}$ is related to the limiting behavior of $\Psi_{e}({\bf r})$
for $r$ approaches zero as
\begin{equation}
\label{2-C}
{\cal C} = \left[\frac{\partial}{\partial r} r \Psi_{e}({\bf r}) \right]_{r \to 0}
\end{equation}

Substituting the solution (\ref{2-coeff}) for coefficients $c_{\bf n}$ into
Eq. (\ref{2-C}),  we can cancel out the constant ${\cal C}$ and obtain
and equation which determines the eigenenergy $E$,
\begin{equation}
\label{2-as}
-\frac{1}{2\pi a_s} = \left[ \frac{\partial}{\partial r}
r \Psi_{e}({\cal E}, {\bf r}) \right]_{r \to 0},
\end{equation}
where
\begin{equation}
\label{2-psi2}
\Psi_{e}({\cal E}, {\bf r}) = \sum_{(n_x, n_y)'} \sum_{n_z}
\frac{\sqrt{\eta \epsilon}
{\cal P}_{\bf n} (0,0,0)
{\cal P}_{\bf n} (x,y,z) }
{n_x \eta + n_y + n_z \epsilon - {\cal E}},
\end{equation}
and ${\cal E} = E - E_0$ denotes the energy shifted by the zero point energy
$E_0 =(\eta + 1+ \epsilon)/2$. It should be emphasized that the regularization operator
$\partial_r r$ and the summation in Eqs. (\ref{2-as}) and (\ref{2-psi2}) cannot be
interchanged, since the series is not uniformly convergent. As we can see later,
this is related to the divergent behavior of $\Psi_{e}$ at small $r$, which is
regularized by $\partial_r (r \cdot)$. To perform the summation in Eq. (\ref{2-psi2}) we
express the denominator in terms of the following integral\cite{Idziaszek-06}
\begin{equation}
\label{2-integral}
\frac{1}{n_x \eta + n_y + n_z \epsilon - {\cal E}}
=
\int_0^\infty dt e^{-t(n_x \eta + n_y + n_z \epsilon - {\cal E})}.
\end{equation}
Notice that the summation in $\Psi_e$ runs over the excited modes with
$n_x + n_y >0$, hence this integral representation is valid for ${\cal E} < \min[\eta, 1]$,
i.e., for the eigenstate with energy less than $E_0 + \min[\eta, 1]$.
Notice that, in the limiting case of $\epsilon \to 0$, the zero point energy
represents the open channel threshold $E_0 (\epsilon \to 0) = E_{g0}$.
Thus, we can safely use the integral representation (\ref{2-integral}) for
${\cal E}  \le 0$, and determine the Feshbach resonance condition.
\begin{figure*}[ht]
\begin{center}
\psfrag{invas}{$a_y/a_s$}
\psfrag{E}{$E(\hbar \omega_y)$}
\psfrag{Eeb}{$E_{eb}$}
\psfrag{Eb}{$E_{b}$}
\psfrag{Ee0}{$E_{e0}$}
\psfrag{Ec0}{$E_{g0}$}
\psfrag{etaone}{(a) $\eta = 1$}
\psfrag{etafive}{(b) $\eta = 5$}
\psfrag{etaten}{(c) $\eta = 10$}
\psfrag{etatwenty}{(d) $\eta = 20$}
\includegraphics[width=8.0cm]{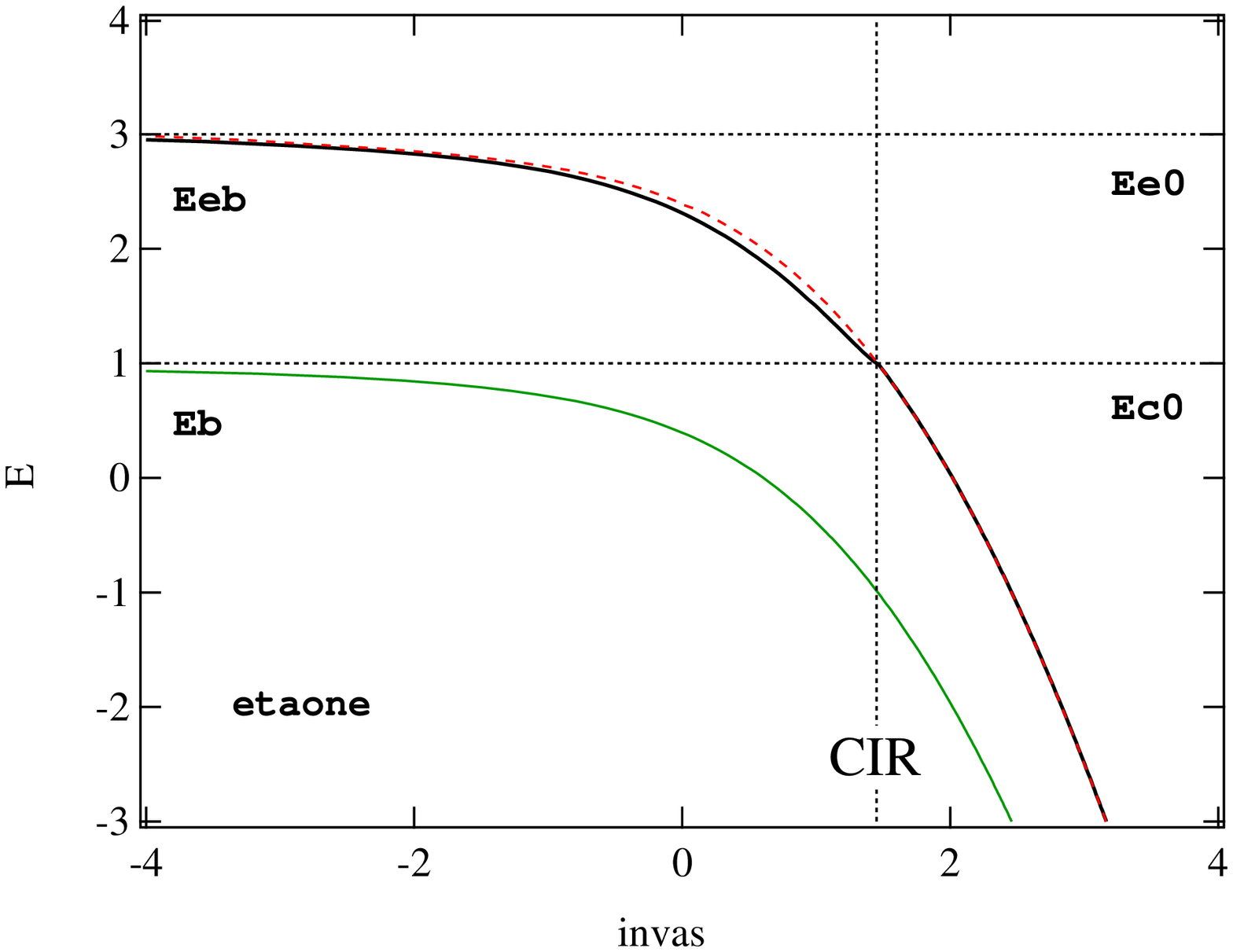}
\hspace{0.5cm}
\includegraphics[width=8.0cm]{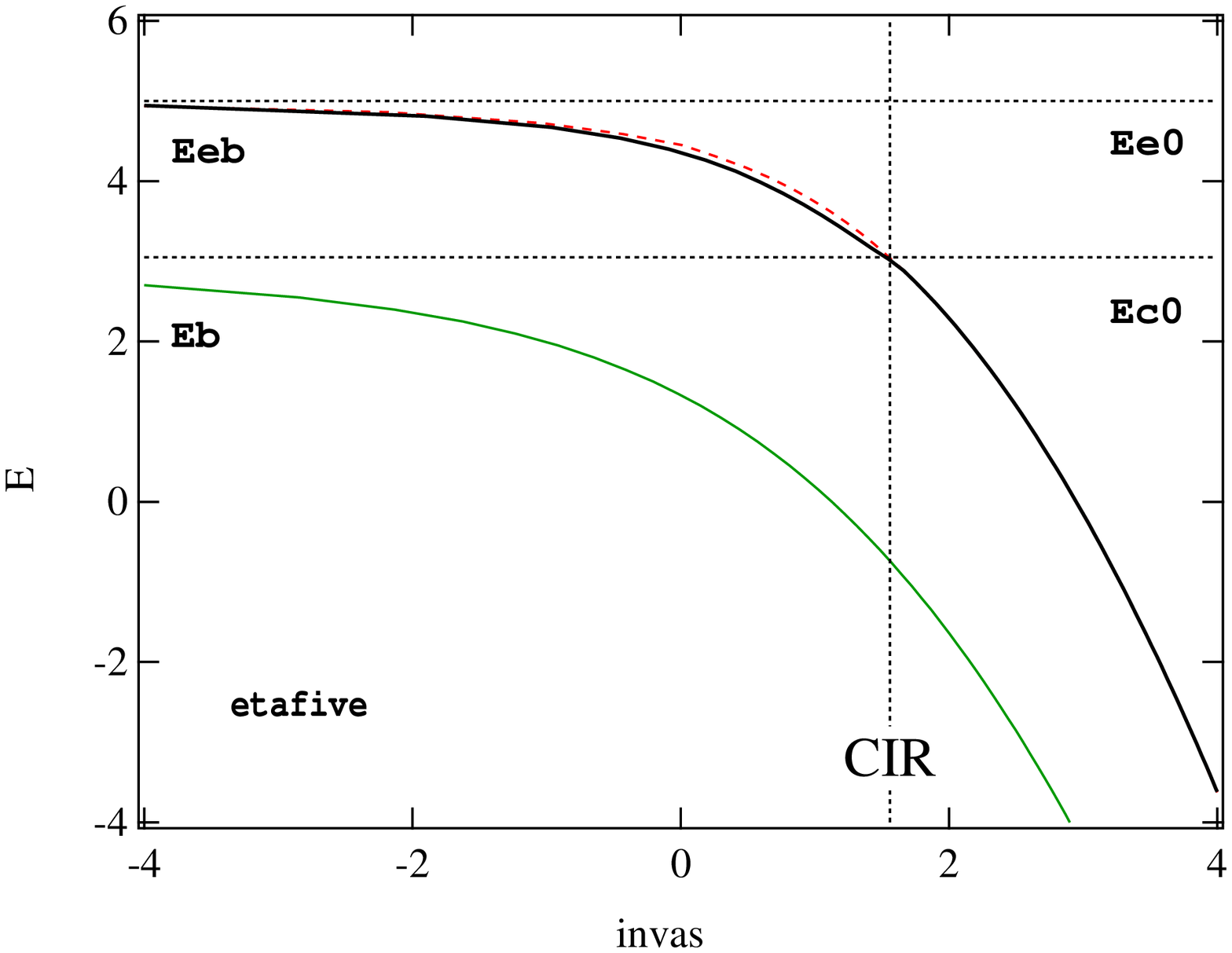}
\\
\includegraphics[width=8.1cm]{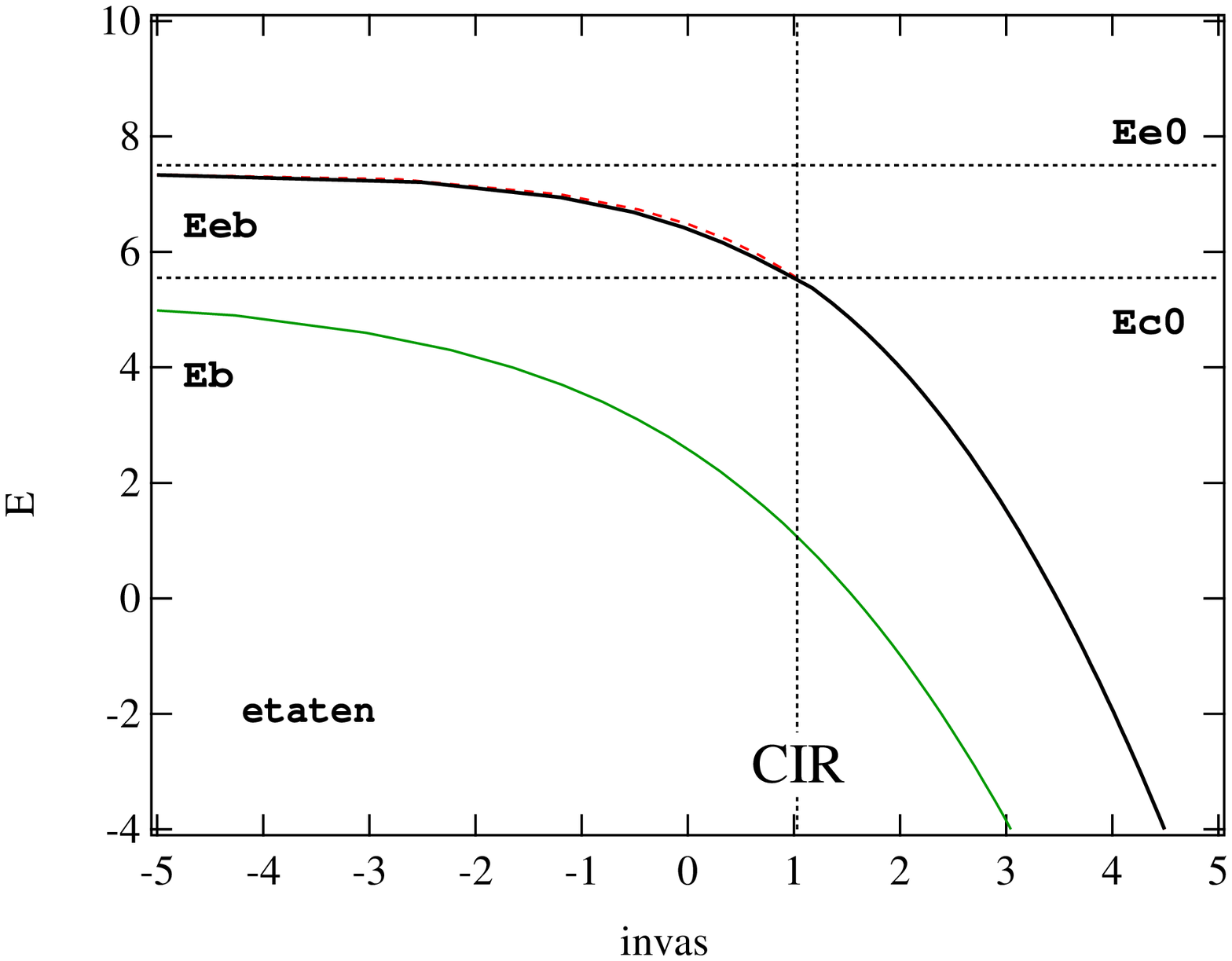}
\hspace{0.3cm}
\includegraphics[width=8.1cm]{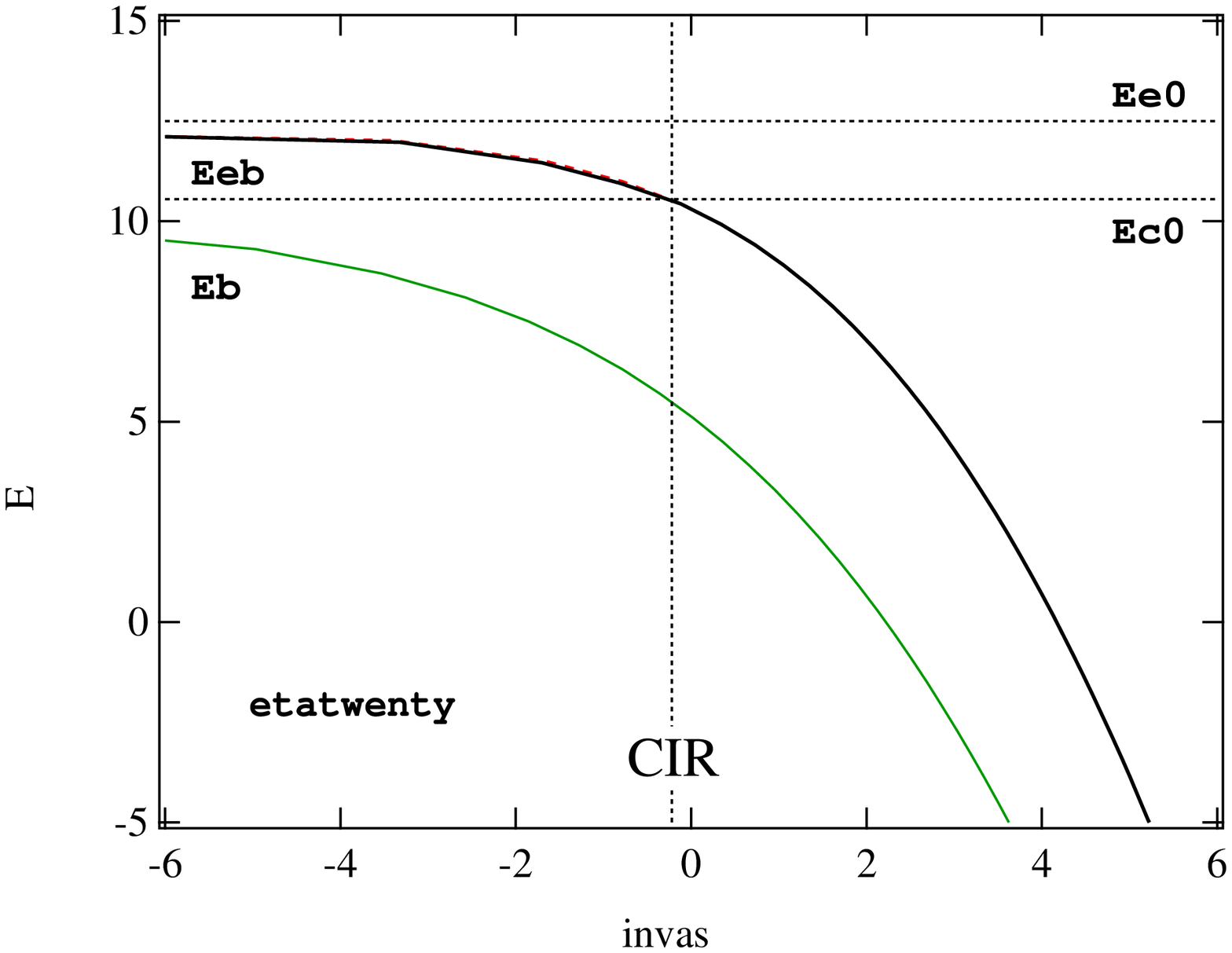}
\caption{(Color online) Bound state energy $E_{\rm eb}$ of the closed channel
Hamiltonian and bound state energy $E_{\rm b}$ of the full Hamiltonian, as
calculated for the quasi-1D limit with $\epsilon = \omega_z /\omega_y \to 0$ (solid).
The results of $E_{\rm eb}$ for small but finite $\epsilon = 0.1$ are also shown
as comparison (dashed). The CIR occurs at the position where $E_{\rm eb}$
coincides with the threshold of the ground Hamiltonian $E_{\rm g0}$, as
schematically drawn in the figures (dotted). }
\label{fig:energy}
\end{center}
\end{figure*}

To perform the summation over $n_x$, $n_y$, and $n_z$, we apply the
generating function for the products of Hermite polynomials~\cite{Prudnikov-86}
\begin{equation}
\label{2-generating}
\sum_{n=0}^\infty \frac{\sigma^n}{2^n n!} H_n(x) H_n(y)
= \frac{e^{(2\sigma x y - \sigma^2 x^2 - \sigma^2 y^2)/(1-\sigma^2)}}
{\sqrt{1-\sigma^2}}.
\end{equation}
Upon inserting Eq. (\ref{2-integral}) into (\ref{2-psi2}), and performing the summation
according to (\ref{2-generating}), we obtain the following integral representation
for the wave function
\begin{equation}
\label{2-psi3}
\Psi_{e}({\cal E},{\bf r}) = \psi_{e}^{\rm I}({\cal E},{\bf r})
+ \psi_{e}^{\rm II} ({\cal E},{\bf r}),
\end{equation}
where
\begin{widetext}
\begin{eqnarray}
\psi_{e}^{\rm I}({\cal E},{\bf r}) &=&
\frac{\sqrt{\eta \epsilon}}{(2\pi)^{3/2}}
\int_0^\infty dt
\frac{e^{t({\cal E}+E_0)}}
{\sqrt{\sinh(\eta t) \sinh(t) \sinh(\epsilon t)}}
\exp\left[- \frac{\eta x^2}{2} \coth(\eta t)- \frac{y^2}{2} \coth( t)
- \frac{\epsilon z^2}{2} \coth(\epsilon t)\right],
\label{2-psiI}
\\
\psi_{e}^{\rm II}({\cal E},{\bf r}) &=&
\frac{-2 \sqrt{\eta \epsilon}}{(2\pi)^{3/2}}
\int_0^\infty dt
\frac{1}{\sqrt{\sinh(t) }}
\exp\left[t \left({\cal E} + \frac{\epsilon}{2} \right) - \frac{\eta x^2 + y^2}{2}
- \frac{\epsilon z^2}{2} \coth(\epsilon t) \right].
\label{2-psiII}
\end{eqnarray}
\end{widetext}

Next we investigate the behavior of the wave function for small values of $r$.
Note that the integral of $\psi_{e}^{\rm II}$ is well behaved at $r = 0$
and can be carried out analytically, leading to
\begin{eqnarray}
\label{2-psiII2}
\psi_{e}^{\rm II} ({\cal E}, r=0) &=&
- \frac{\sqrt{\eta \epsilon}}{\pi^{3/2}}
\int_0^\infty dt
\frac{e^{t {\cal E}}}
{\sqrt{1-e^{-2 \epsilon t}}}
\nonumber \\
&=&
-\frac{\sqrt{\eta}}{2\pi \sqrt{\epsilon}}
\frac{\Gamma(- {\cal E}/2 \epsilon)}{\Gamma(- {\cal E}/2 \epsilon + 1/2)}.
\end{eqnarray}
On the other hand, $\psi_{e}^{\rm I}$ is divergent in the limit of $r \to 0$,
where the main contribution to the integral comes from small arguments $t$.
In the leading order we neglect the energy dependence of ${\cal E}$,
and the expansion of (\ref{2-psiI}) for small $t$ reads
\begin{equation}
\label{2-psiI2}
\psi_{e}^{\rm I} ({\cal E}, {\bf r}) \approx
\frac{1}{(2\pi)^{3/2}}
\int_0^\infty dt \frac{e^{-r^2/(2t)}}{t^{3/2}}
=
\frac{1}{2 \pi r}.
\end{equation}
By adding the two terms together, we note that the wave function $\Psi_{e}$
diverges in the same way as the Green's function in a homogeneous space.
This can be understood by reminding that at short distances the behavior of
the wave function is mainly determined by the interaction between particles,
and has negligible dependence on the global trapping potential.
Substituting the expression of wave function $\Psi_{e}$ into (\ref{2-as}),
the divergent behavior will be regularized by the operator $\partial_r (r \cdot)$,
hence make no contribution to the equation. Therefore, we can subtract from
$\Psi_{e}$ the right-hand side of (\ref{2-psiI2}), and obtain a simpler
expression determining the bound state energy,
\begin{equation}
\label{2-as2}
\frac{-\sqrt{\pi}}{a_s} = {\cal F}({\cal E}),
\end{equation}
where
\begin{eqnarray}
\label{2-F}
{\cal F}({\cal E}) &=& \int_0^\infty dt \bigg[
\frac{\sqrt{\eta \epsilon} e^{t {\cal E} /2}}
{\sqrt{(1-e^{-\eta t}) (1- e^{-t}) (1- e^{-\epsilon t})}}
\nonumber \\
&& \hspace{2cm}
- \frac{\sqrt{\eta \epsilon} e^{t {\cal E}/2}}{\sqrt{1-e^{-\epsilon t}}}
- \frac{1}{t^{3/2}}
\bigg].
\end{eqnarray}
Notice that this representation is valid for all values of ${\cal E} \le 0$.
In the limiting case of ${\cal E} =0$, the second and the third terms remove
the divergence for $t \to \infty$ and $t \to 0$, respectively, hence
guarantee the convergence of ${\cal F}(0)$.

Finally, we turn to the discussion of quasi-1D regime by pushing to the
limit of $\epsilon \to 0$. This corresponds to lowering the axial ($z$) trapping
frequency such that it is much weaker than the transverse $x$-$y$ confinement.
In this limiting case, the main contribution to the integral of ${\cal F}$ comes from
small arguments $t$, and we perform the approximation
$\sqrt{1-e^{-\epsilon t}} \approx \sqrt{\epsilon t}$ in the determinator to obtain
\begin{eqnarray}
\label{2-F2}
{\cal F}({\cal E}) &\approx& \int_0^\infty dt
\bigg[
\frac{\sqrt{\eta} e^{t {\cal E}/2}}
{\sqrt{t (1-e^{-\eta t}) (1- e^{-t}) }}
\nonumber \\
&& \hspace{2cm}
-
\frac{\sqrt{\eta} e^{t {\cal E}/2}}
{\sqrt{t}}
-
\frac{1}{t^{3/2}}
\bigg].
\end{eqnarray}
Considering the fact that the CIR takes place when the eigenenergy
$E_{eb} = {\cal E} + E_0 = E_{g0}$, we reach the equation for the
CIR condition in physical units as
\begin{equation}
\label{2-as3}
\frac{a_s}{a_y} = -\frac{\sqrt{\pi}}{{\cal F}(0)}.
\end{equation}
A numerical evaluation of function ${\cal F}$ shows that this condition
predicts exactly the same CIR position as (\ref{1-CIR}), as one should expect.
In particular, in the cylindrically symmetric case of $\eta =1$, the integral
of ${\cal F}$ can be performed analytically, leading to
\begin{eqnarray}
\label{2-F3}
{\cal F}({\cal E}) &=& \int_0^\infty dt
\left[
\frac{e^{-t(1-{\cal E}/2)}}
{\sqrt{t} (1- e^{-t})}
- \frac{1}{t^{3/2}}
\right]
\nonumber \\
&=&
\sqrt{\pi}
\zeta_{\rm H} \left( \frac{1}{2}, 1 - \frac{\cal E}{2} \right),
\end{eqnarray}
where $\zeta_{\rm H} (s,a) \equiv \sum_{n=0}^\infty (n+a)^{-s}$
denotes the Hurwitz Zeta function. Upon substituting the expression above
into the CIR condition (\ref{2-as3}), and noticing that
$\zeta_{\rm H} (1/2, 1) = \zeta(1/2)$, we recover the known result
of (\ref{1-CIR4}).

In Fig. 3 we show the bound state energy $E_{eb}$ of the closed channel Hamiltonian
as a function of 3D scattering length $a_s /a_y$ for several different choices of $\eta$.
In each plot, the location of CIR is indicated by dotted lines as the eigenenergy
crosses the open channel threshold $E_{g0}$. We also verify that the quasi-1D
regime for two interacting atoms does not require an infinitesimally small value of $\epsilon$,
but is already realized for $\epsilon \sim 0.1$ as illustrated in Fig. 3 as (red) dashed lines.
We have also studied the case of $\epsilon = 0.001$, which is close to the
realistic experimental setup~\cite{Haller-10}, and note an excellent agreement with
the quasi-1D energy levels, which on the scale of Fig. 3 are indistinguishable.

It is also illuminating to investigate the bound state energy of the full Hamiltonian
as ${\hat H}_{\rm rel} \Psi({\bf r}) = E_{b} \Psi({\bf r})$. This can be done by
decomposing the wave function into the complete set of harmonic oscillators
\begin{equation}
\label{2-Psi4}
\Psi_e ({\bf r}) = \sum_{n_x,n_y, n_z} c_{n_x n_y n_z}
\Phi_{n_x}(\sqrt{\eta} x) \Phi_{n_y} (y) \Phi_{n_z} (\sqrt{\epsilon} z),
\end{equation}
and following the same procedure as in derivation of (\ref{2-F}). As a result,
the bound state energy is determined by the equation
\begin{eqnarray}
\label{2-eqn2}
&&\int_0^\infty dt \bigg[
\frac{\sqrt{\eta \epsilon} e^{t (E_{b} - E_0) /2}}
{\sqrt{(1-e^{-\eta t}) (1- e^{-t}) (1- e^{-\epsilon t})}}
\nonumber \\
&& \hspace{2.5cm}
-
\frac{1}{t^{3/2}}
\bigg] =
\frac{-\sqrt{\pi}}{a_s}.
\end{eqnarray}
Upon comparing with the equation for the bound state energy of the
closed channel Hamiltonian (\ref{2-eqn}), we notice that in the
cylindrically symmetric case of $\eta =1$, the bound state energy
$E_{eb}$ of the closed channel Hamiltonian and the bound
state energy $E_{b}$ of the full Hamiltonian is related via
\begin{equation}
E_{eb} = E_{b} + 2,\quad \quad \textrm{for } \eta =1.
\end{equation}
For general cases of arbitrary $\eta$, the difference between $E_{eb}$
and $E_{b}$ is in general has dependence of $a_s$, as shown
in Fig. 3 (green/gray solid lines).


\section{The two-channel calculation for general forms of interaction}
\label{sec:formal theory}

In the previous discussion of CIR, we have employed a single-channel
zero range pseudo-potential to model the 3D interatomic interaction.
Nevertheless, in the realistic cases, the atom-atom interactions have
finite ranges and multi channels with respect to different hyperfine states.
The 3D scattering length $a_s$ is determined by the inter-channel
Feshbach resonance, which is controlled by magnetic field $B$ through
the $B$-dependentinter-channel detuning.
To take into account of these characters, in this section we provide a
formal calculation of the effective quasi-1D scattering length $a_{\mathrm{1D}}$
with a general two-channel model which is more realistic for the interatomic
interaction. Our formal calculation can explicitly give the qualitative
behavior of $a_{\mathrm{1D}}$ with respect to the magnetic field $B$, and
show clearly that there is only {\it one} resonant point where $a_{\mathrm{1D}}$
takes zero value. This is consistent with our above results based on the
usage of pseudo-potential.

In the two-channel model, the atoms are initially localized in the open
channel with a two-atom hyperfine state $|o\rangle _{s}$. During the
scattering process, the atomic motion in the open channel is assumed to be
near resonant with a bound state $|\Phi \rangle $ in the closed channel, whose
threshold energy is much higher than the scattering energy. The
inter-channel coupling can induce the quantum transition between the two
channels and then vary the scattering amplitude.

In our calculation, the Hamiltonian for the atomic relative motion takes the form
\begin{eqnarray}
{\hat H}_{\rm rel}={\hat T}_{z} + {\hat H}_{\perp }
+\varepsilon \left( B\right) |c\rangle _{s}\langle c|+{\hat V}.
\label{bigh}
\end{eqnarray}
where $\varepsilon \left( B\right) $ is the threshold energy of the closed
channel with respect to the hyperfine state $|c\rangle _{s}$, and can be tuned
as a linear function of the magnetic field $B$.
In this Hamiltonian, ${\hat T}_{z}$ is the kinetic energy in the $z$ direction,
${\hat H}_{\bot }$ is the Hamiltonian of the transverse degrees of freedom.
As in the above sections, we denote the ground state of the transverse Hamiltonian
${\hat H}_{\perp }$ as $|00\rangle_\perp$. The two-channel interaction term ${\hat V}$
can be further expressed as
\begin{eqnarray}
{\hat V} &=& V_{\mathrm{bg}} \left( r\right) |o\rangle _{s}\langle o|
+V_{\mathrm{cl}}\left( r\right) |c\rangle _{s}\langle c|
\nonumber \\
&& \hspace{1cm}
+ \left[ W\left( r\right) |o\rangle_{s}\langle c|+h.c. \right],
\end{eqnarray}
where $r$ is the 3D relative coordinate between the two atoms,
$V_{\mathrm{bg}}\left( r\right) $ and $V_{\mathrm{cl}}\left( r\right) $ are
the interatomic interaction potentials in the open and closed channels, respectively,
$W\left( r\right) $ the inter-channel coupling, and $h.c.$ denotes Hermitian conjugate.
For simplicity, we assume $V_{\mathrm{bg}}\left( r\right) $ is a short-range
potential with
\begin{eqnarray}
V_{\mathrm{bg}}\left( r\right)=0,\ r>r_{\ast} \label{simple}
\end{eqnarray}
Such a simple model is reasonable for the formal calculation
if the realistic atom-atom interaction potential
is a short-range potential (e.g., the Yukawa potential) which decays
faster than any power-law function of the atom-atom distance $r$.
In the ultracold gases, the atoms experience a long-range
van der Waals potential which behaves
as $1/r^6$ in the long-range region, rather than a short-range one.
However, it has been proved that in the low-energy limit
there is no difference between the behaviors of the 3D $s$-wave scattering amplitude from the
van der Waals potential and the short-range potential~\cite{JMP1,gao1}. Therefore,
we use Eq.(\ref{simple}) as a reasonable simplification
for the discussion on the low-energy scattering.

In our system the spatial parity along the $z$ direction is conserved. Since
the $s$-wave bound state in the closed channel has an even parity, the
Feshbach resonance between the open and closed channels can only change
the scattering amplitude for the states with even parity in the open channel.
Namely, the scattering amplitude with respect to odd parity is independent
on the magnetic field $B$. Therefore, in this section we only concentrate on the
scattering amplitude for the states with even parity.

The 1D effective scattering length $a_{\mathrm{1D}}$ can be evaluated
from the analysis of scattering states. Before the scattering process, the
two atoms are moving in the ground state $|00\rangle_\perp$ of the transverse
Hamiltonian ${\hat H}_{\perp }$ with relative momentum $k$. Then the
relevant scattering state $|\psi ^{\left(+\right) }\left( k\right) \rangle $ is given
by the Lippmman-Schwinger equation~\cite{taylor}
\begin{eqnarray}
|\psi ^{\left( +\right) }\left( k\right) \rangle
=
|k\rangle_{z} |00\rangle_{\bot } |o\rangle_{s}
+ G_{0}^{(+)}(k^{2}) {\hat V} |\psi ^{\left( + \right) } \left( k\right) \rangle
\label{OLP}
\end{eqnarray}
where $|k\rangle _{z}$ is the eigenstate of ${\hat T}_{z}$ with even parity
$_{z}\langle z|k\rangle _{z}=\cos kz$, and $G_{0}^{(+)}(k^{2})$ is
the free Green's operator
\begin{eqnarray}
G_{0}^{(+)}(k^{2}) =
\frac{1}{k^{2}+i0^{+} - {\hat T}_{z} - {\hat H}_{\perp } -
\varepsilon \left( B \right) |c\rangle_{s} \langle c|}.
\end{eqnarray}
Throughout this section, we use the natural unit with $\hbar = \mu = 1$.
The state $|\psi ^{\left( +\right) }\left( k\right) \rangle $ can be
rewritten as
\begin{eqnarray}
|\psi ^{\left( +\right) }\left( k\right) \rangle =|\psi _{o}\left( k\right)
\rangle |o\rangle _{s}+|\psi _{c}\left( k\right) \rangle |c\rangle _{s},
\label{rpsi}
\end{eqnarray}
where the states $|\psi _{o,c}\left( k\right) \rangle $ are the spatial
states correspond to the open and closed channels, respectively.
As shown in Appendix \ref{sec-appendix-A}, the Lippmman-Schwinger equation (\ref{OLP})
of $|\psi ^{\left( +\right)}\left( k\right) \rangle $ can be re-expressed as~\cite{julian}
\begin{eqnarray}
|\psi _{o}\left( k\right) \rangle &=&|\psi _{\mathrm{bg}}^{\left( +\right)
}\left( k\right) \rangle +G_{\mathrm{bg}}^{\left( +\right) }\left( k\right)
W|\psi _{c}\left( k\right) \rangle,  \label{LPe1} \\
|\psi _{c}\left( k\right) \rangle &=&\frac{1}{k^{2}+\Delta \left( B\right) }
|\Phi \rangle \langle \Phi |W|\psi _{o}\left( k\right) \rangle,  \label{LPe2}
\end{eqnarray}
where $|\psi_{\mathrm{bg}}^{\left( +\right) }\left( k\right) \rangle $ is
the background scattering state defined by the\ background
Lippmman-Schwinger equation
\begin{eqnarray}
|\psi _{\mathrm{bg}}^{\left( +\right) }\left( k\right) \rangle =|k\rangle
_{z}|00\rangle _{\bot }+G_{\mathrm{bg}}^{\left( +\right) }\left( k\right)
V_{\mathrm{bg}}\left( r\right) |k\rangle _{z}|00\rangle_{\bot }.  \label{LPbg}
\end{eqnarray}
Here, $G_{\mathrm{bg}}^{\left( +\right) }\left( k\right) $ is the background
Green's operator given by
\begin{eqnarray}
G_{\mathrm{bg}}^{\left( +\right) }\left( k\right) =
\frac{1}{k^{2}+i0^{+} - {\hat T}_{z} + {\hat H}_{\perp }+V_{\mathrm{bg}}}.  \label{ggbg}
\end{eqnarray}
In Eq. (\ref{LPe2}) the detuning $\Delta \left( B\right) $ is defined as
\begin{eqnarray}
\Delta \left( B\right) =E_{b}-\varepsilon \left( B\right),  \label{deltab}
\end{eqnarray}
where $E_{b}$ is the binding energy of the bound state $|\Phi \rangle $ in
the closed channel.

Substituting Eq. (\ref{LPe2}) into (\ref{LPe1}), and multiplying the
Dirac bra $\langle \Phi |W$ on both the left-hand and right-hand sides of (\ref{LPe1}),
we obtain a linear equation for the factor $\langle \Phi |W|\psi_{o}\left( k\right) \rangle $.
The solution of this equation leads the open-channel component of the scattering state
\begin{eqnarray}
|\psi _{o}\left( k\right) \rangle =|\psi _{\mathrm{bg}}^{\left( +\right)}\left( k\right) \rangle
+
\frac{G_{\mathrm{bg}}^{\left( +\right) }\left(
k\right) W|\Phi \rangle \langle \Phi |W|\psi _{\mathrm{bg}}^{\left( +\right)
}\left( k\right) \rangle }{\Delta \left( B\right)
-
\langle \Phi |WG_{\mathrm{bg}}^{\left( +\right) }\left( k\right) W|\Phi \rangle }.
\notag \\
\end{eqnarray}

The 1D scattering length $a_{\mathrm{1D}}\left( B\right) $ can be extracted from
the asymptotic behavior of the open-channel spatial state
$|\psi _{o}\left( k\right) \rangle $ in the long-distance limit
$|z|\longrightarrow\infty$. As shown in Appendix \ref{sec-appendix-D},
in this region we have
\begin{eqnarray}
_{z}\langle z|\psi _{\mathrm{bg}}^{\left( +\right) }\left( k\right) \rangle
&=&
\left( \cos kz+f_{\mathrm{even}}^{\mathrm{bg}}\left( k\right)
e^{ik\left\vert z\right\vert }\right) |00\rangle _{\perp },
\\
_{z}\langle z|G_{\mathrm{bg}}^{\left( +\right) }\left( k\right) &=&
-i \frac{e^{ik\left\vert z\right\vert }}{2k}|00\rangle _{\perp }
\langle \psi _{\mathrm{bg}}^{\left( -\right) }\left( k\right) |,
\label{asygbg}
\end{eqnarray}
where $|\psi _{\mathrm{bg}}^{\left( -\right) }\left( k\right) \rangle $ is
the background scattering state with in-going boundary condition and
$f_{\mathrm{even}}^{\mathrm{bg}}\left( k\right) $ is the background
scattering amplitude with low-energy behavior
\begin{eqnarray}
f_{\mathrm{even}}^{\mathrm{bg}}\left( k\to 0\right) =
-1 + ika_{\mathrm{1D}}^{\mathrm{bg}}.
\end{eqnarray}
Here, $a_{\mathrm{1D}}^{\mathrm{bg}} \in \mathrm{real}$ is the background
scattering length.

We can also prove that $G_{\mathrm{bg}}^{\left(+\right) }\left( k=0\right) $
is a Hermitian operator as shown in Appendix \ref{sec-appendix-D}. Therefore, in the
long-distance region we have
\begin{eqnarray}
_{z}\langle z|\psi _{o}^{\left( +\right) }\left( k\right) \rangle =\left(
\cos kz+f_{\mathrm{even}}\left( k;B\right) e^{ik\left\vert z\right\vert
}\right) |00\rangle_{\perp }
\end{eqnarray}
with the scattering amplitude given by
\begin{eqnarray}
f_{\mathrm{even}}\left( k\to 0;B\right)
=
-1+ika_{\mathrm{1D}}\left( B\right)
\end{eqnarray}
in the low-energy limit. Here, the $B$-dependent scattering length
$a_{\mathrm{1D}}\left( B\right) $ takes the form
\begin{eqnarray}  \label{ra1d}
a_{\mathrm{1D}}\left( B\right) =a_{\mathrm{1D}}^{\mathrm{bg}%
}+\lim_{k\to 0}\frac{1}{2k^{2}}\frac{\left\vert \langle \Phi |W|\psi
_{\mathrm{bg}}^{\left( +\right) }\left( k\right) \rangle \right\vert ^{2}}{%
\Delta(B) -\langle \Phi |WG_{\mathrm{bg}}^{\left( +\right) }\left(0\right)
W|\Phi \rangle }.  \notag \\
\end{eqnarray}
To obtain the expression (\ref{ra1d}), we have also used the results
$|\psi _{\mathrm{bg}}^{(+) }( k) \rangle \propto k$ and
$|\psi _{\mathrm{bg}}^{(-) }(k) \rangle =-|\psi _{\mathrm{bg}}^{(+) }( k) \rangle$,
which are proved in the limiting case of $k\to 0$ as shown in Appendix \ref{sec-appendix-D}.

The result of $a_{\rm 1D}(B)$ expressed in (\ref{ra1d}) is the central result
of this section. It gives the behavior of $a_{\mathrm{1D}}$ as a function of
the magnetic field $B$ through the magnetic Feshbach resonance.
Notice that $\Delta(B)$ is a linear function of $B$, thus one can immediately read
from Eq. (\ref{ra1d}) that there is only {\it one} resonant point
\begin{eqnarray}
B_{\ast}=\frac{1}{\mu}\left(E_b-\langle\Phi|WG_{\rm bg}^{+}(0)W|\Phi\rangle+\Delta_a/a_{\rm 1D}^{\rm bg}\right)
\end{eqnarray}
leading to zero value of $a_{\mathrm{1D}}\left( B_{\ast }\right) =0$. Here we have
assumed $\varepsilon(B)=\mu B$ and $\Delta_a$ is defined as
\begin{eqnarray}
\Delta_a=\lim_{k\to 0}\frac{1}{2k^{2}}\left\vert \langle \Phi |W|\psi
_{\mathrm{bg}}^{\left( +\right) }\left( k\right) \rangle \right\vert ^{2}.
\end{eqnarray}
This observation is in consistent with our findings in the previous
sections by assuming a $\delta$ pseudo-potential.
We stress that this formal scattering calculation is based on a
very general form of the quasi-1D Hamiltonian (\ref{bigh}), and the
qualitative conclusion of one single CIR remains valid disregard
to the details of transverse confinement or the interatomic potential.


\section{Conclusion}
\label{sec:conclusion}

In summary, we have presented a detailed analysis of the scattering process
of two interacting atoms confined in a quasi-1D harmonic trap. We analyzed
the possibility of confinement-induced resonance (CIR), where the 1D
scattering degenerates to a total reflection as a 1D gas of impenetrable bosons.
By modeling the interatomic interaction by a 3D $\delta$ pseudo-potential,
we have shown that only one {\it single} CIR would be present in a quasi-1D trap
with arbitrary transverse anisotropy. The location of this resonance has
non-monotonic dependence on the transverse anisotropy ratio
$\eta = \omega_x / \omega_y$. For slight transverse anisotropy with $\eta \sim 1$,
the position of CIR varies with the same trend as the experimental observation
of CIR2~\cite{Haller-10}, but deviates from the measurements in a quantitative level.
The other resonance observed in experiment, CIR1, cannot be accounted
for the present theory. To extend the discussion to general forms
of $s$-wave interaction, we presented a formal theory within a two-channel model,
and draw the same qualitative conclusion of one CIR for transverse confinement
with arbitrary anisotropy.

After finalizing this manuscript, we are aware of a work recently posted
by Peng {\it et al.}~\cite{Peng-10}, which discusses CIRs in quasi-1D geometries with
transverse anisotropy and reaches similar results as our calculation with pseudo-potential.
However, the formal theory which is valid for a more general form of interaction is not
discussed in~\cite{Peng-10}.


\acknowledgments

We would like to thank L. You for discussion.
This work is supported by NSFC (10904172, 32510018), the Fundamental Research Funds
for the Central Universities, and the Research Funds of Renmin University of
China (10XNF033, 10XNL016). WZ would also like to thank China Postdoctoral
Science Foundation for support.


\appendix

\section{Background Lippmman-Schwinger Equation}
\label{sec-appendix-A}

In this appendix we derive the Lippmman-Schwinger equations (\ref{LPe1}) and
(\ref{LPe2}) for the scattering state $|\psi ^{\left( +\right) }\left(
k\right) \rangle $. We first substitute Eq. (\ref{rpsi}) into (\ref{OLP}),
and get
\begin{eqnarray}
|\psi _{o}\left( k\right) \rangle &=&|k\rangle _{z}|00\rangle _{\bot
}+G_{\rm fo}^{(+)}(k^{2})V_{\mathrm{bg}}|\psi _{o}\left( k\right) \rangle  \notag
\\
&&+G_{\rm fo}^{(+)}(k^{2})W|\psi _{c}\left( k\right) \rangle  \label{lp1a} \\
|\psi _{c}\left( k\right) \rangle
&=&G_{\rm fc}(k^{2})V_{\mathrm{cl}}|\psi_{c}\left( k\right) \rangle
+G_{\rm fc}(k^{2})W|\psi _{o}\left( k\right) \rangle
\notag \\
\label{lp1b}
\end{eqnarray}
with the free Green's operators $G_{\rm fo}^{(+)}$ and $G_{\rm fc}$ in the
open can closed channels, respectively,
\begin{eqnarray}
G_{\rm fo}^{(+)}(k^{2}) &=&\frac{1}{k^2+i0^{+}-{\hat T}_z-{\hat H}_{\perp}},
\\
G_{\rm fc}(k^{2}) &=&\frac{1}{k^2-{\hat T}_z-{\hat H}_{\perp}-\varepsilon (B)}.
\end{eqnarray}
By defining the closed channel Green's function $G_{\mathrm{cl}}$
\begin{eqnarray}
G_{\mathrm{cl}}(k^{2})=\frac{1}{k^{2}-{\hat T}_{z}-{\hat H}_{\perp }-V_{\mathrm{cl}}
- \varepsilon (B)},
\label{gcl}
\end{eqnarray}
we obtain the relations
\begin{eqnarray}
\label{g01}
G_{\rm fo}^{(+)}(k^{2}) &=&G_{\mathrm{bg}}^{(+)}(k^{2})
- G_{\mathrm{bg}}^{(+)}(k^{2})
V^{(\mathrm{bg})}
G_{\rm fo}^{(+)}(k^{2});  \notag \\
\\
G_{\rm fc}(k^{2}) &=&G_{\mathrm{cl}}(k^{2})
-
G_{\mathrm{cl}}(k^{2})V^{(\mathrm{cl})}G_{\rm fc}(k^{2}),
\label{g02}
\end{eqnarray}
where $G_{\mathrm{bg}}^{(+)}(k^{2})$ is defined in (\ref{ggbg}).
Substituting Eqs. (\ref{g01}) and (\ref{g02}) into the last terms of the
right-hand-side (r.h.s.) of Eqs. (\ref{lp1a}) and (\ref{lp1b}), and using
the Lippmman-Schwinger equation (\ref{LPbg}) for the background scattering
state $|\psi _{\mathrm{bg}}^{\left( +\right) }\left( k\right) \rangle $ as
well as the approximation%
\begin{eqnarray}
G_{\mathrm{cl}}(k^{2})\approx \frac{|\Phi \rangle \langle \Phi |}{k^2+\Delta
\left( B\right) }
\end{eqnarray}%
with $\Delta(B)$ defined in Eq. (\ref{deltab}), we get equations (\ref%
{LPe1}) and (\ref{LPe2}) for the states $|\psi _{o}\left( k\right) \rangle $
and $|\psi _{c}\left( k\right) \rangle $.


\section{Low-energy Behavior of 1D Scattering State and Green's Operator}
\label{sec-appendix-B}

In this appendix we discuss the low-energy behavior of the scattering state
in a single-channel 1D scattering problem. The results of this discussion are not
directly used in the main text of our paper, but they are prerequisite for the calculation
in Appendix \ref{sec-appendix-D}.

We consider a 1D scattering problem with symmetric short-range potential with
\begin{eqnarray}
U\left( z\right)  &=&U\left( -z\right) ; \\
U\left( \left\vert z\right\vert >r_{\ast }\right)  &=&0.
\end{eqnarray}
The wave function $\phi ^{\left( +\right) }\left( k;z\right) $ of the
scattering state with even parity hence satisfies the Lippmman-Schwinger
equation
\begin{eqnarray}
\phi ^{\left( +\right) }\left( k;z\right)  &=&\cos kz+
\int dz^{\prime} \Big[
g_{0}^{\left( +\right) }\left( k;z,z^{\prime }\right)
\nonumber \\
&& \hspace{1cm}
\times U\left( z^{\prime}\right)
\phi ^{\left( +\right) }\left( k;z\right) \Big]
\label{OLPa}
\end{eqnarray}
where $g_{0}^{\left( +\right) }\left( k;z,z^{\prime }\right) $ is the 1D Green's function
\begin{eqnarray}
g_{0}^{\left( +\right) }\left( k;z,z^{\prime }\right)  &=&_{z}\langle z|%
\frac{1}{k^{2}+i0^{+}-{\hat T}_{z}}|z^{\prime }\rangle _{z}  \notag \\
&=&-i\frac{1}{2k}e^{ik\left\vert z-z^{\prime }\right\vert }.  \label{g0}
\end{eqnarray}
It is straightforward to show that the Lippmman-Schwinger equation
(\ref{OLPa}) is equivalent with the time-independent Schr\"{o}dinger
equation~\cite{taylor}
\begin{eqnarray}
-\frac{d^{2}}{dz^{2}}\phi ^{\left( +\right) }\left( k;z\right) +U\left(
z\right) \phi ^{\left( +\right) }\left( k;z\right) =k^{2}\phi ^{\left(
+\right) }\left( k;z\right) \label{1de}
\end{eqnarray}
with the boundary condition
\begin{eqnarray}
\phi ^{\left( +\right) }\left( k;\left\vert z\right\vert >r_{\ast }\right)
=\cos kz+f\left( k\right) e^{ik\left\vert z\right\vert }.
\label{bod}
\end{eqnarray}
Here, $f(k)$ is the scattering amplitude.

To investigate the low-energy behavior of $\phi ^{\left( +\right) }\left(k;z\right) $
and $f\left( k\right) $, we expand $\phi ^{\left( +\right)}\left( k;z\right) $ in series of $k$
\begin{eqnarray}
\phi ^{\left( +\right) }\left( k;z\right)  &=&\phi _{0}\left( z\right) +\phi
_{1r}\left( z\right) k+i\phi _{1i}\left( z\right) k+{\cal O}\left( k^{2}\right),
\nonumber\\ \label{psik}
&&
\end{eqnarray}
where $\phi _{1r}\left( z\right) $ and $\phi _{1i}\left( z\right) $ are
assumed to be real functions of $z$. Setting $k=0$ in Eq. (\ref{1de}),
we get the equation for $\phi _{0}\left( z\right) $
\begin{eqnarray}
-\frac{d^{2}}{dz^{2}}\phi _{0}\left( z\right) +U\left( z\right)
\phi_{0}\left( z\right) =0. \label{c1}
\end{eqnarray}
Considering the fact that the potential $U\left( z\right) $ takes zero value
in the region of $\left\vert z\right\vert >r_{\ast }$, the solution $\phi _{0}\left( z\right) $
of the equation above would be a constant within this outer region.
Therefore, in the inner region of $\left\vert z\right\vert \leq r_{\ast }$,
we acquire the following boundary condition
\begin{eqnarray}
\left. \frac{d}{dz}\phi _{0}\left( z\right) \right\vert _{\left\vert
z\right\vert =r_{\ast }}=0. \label{c2}
\end{eqnarray}
This boundary condition is a strong requirement such that
only some special kinds of $U\left( z\right) $
(e.g., the most trivial case with $U\left( z\right) =0$)
can lead to a non-zero solution of Eq. (\ref{c1}). Here we assume our
potential are not of this special form, then we have
\begin{eqnarray}
\phi _{0}\left( z\right) &=&0,
\nonumber \\
\phi ^{\left( +\right) }\left( k\to 0;z\right) &\propto& k.
\end{eqnarray}
Substituting the equation above into (\ref{bod}), we obtain the low-energy
behavior
\begin{eqnarray}
f\left( k\to 0\right) =-1 + {\cal O}(k)
\end{eqnarray}
of the scattering amplitude with even parity.

Now we focus on the linear terms of $k$.
Expanding Eqs. (\ref{1de}) to the first order of $k$, we find that
the functions $\phi _{1r}\left( z\right) $ and $\phi _{1i}\left( z\right) $
defined in Eq. (\ref{psik}) also satisfy Eq. (\ref{c1}). The expansion of Eq. (\ref{bod})
further shows that $\phi_{1r}\left( z\right)$ takes
constant value in the outer region of $\left\vert z\right\vert >r_{\ast }$. Thus,
in the inner region of $\left\vert z\right\vert \leq r_{\ast }$
the wave function $\phi _{1r}\left( z\right) $ satisfies the same equation
and boundary condition as $\phi _{0}\left( z\right) $, leading to
\begin{eqnarray}
\phi _{1r}\left( z\right) =\phi _{0}\left( z\right) =0.
\end{eqnarray}
As a consequence, the scattering state $\phi \left( k;z\right) $ with even
parity acquires the following form in the low-energy limit $k\to 0$
\begin{eqnarray}
\phi ^{\left( +\right) }\left( k\to 0;z\right) =ik\phi _{1i}\left(
z\right) ;
\quad \phi _{1i}\left( z\right) \in \mathrm{real}  \label{phizheng}
\end{eqnarray}
which yields
\begin{eqnarray}
f\left( k\to 0\right) =-1 + ika + {\cal O}(k^2);
\quad a\in \mathrm{real}.
\end{eqnarray}

Next, we discuss the low-energy behavior of the 1D Green's operator
\begin{eqnarray}
g(k^{2})=\frac{1}{k^{2}-{\hat T}_{z}-U\left( z\right) }.
\end{eqnarray}
In the limiting case of $k\to 0$, the Green's operator can be expressed as
\begin{eqnarray}
&&_{z}\langle z|g^{\left( +\right) }(k^{2} \to 0)|z^{\prime }\rangle_{z}
\nonumber \\
&& \hspace{1cm}
=-\int_{0}^{\infty }dp\frac{\phi ^{\left( +\right) \ast }\left(
p;z\right) \phi ^{\left( +\right) }\left( p;z^{\prime }\right) }{p^{2}}.
\notag  \label{ge}
\end{eqnarray}
The convergence of this integral representation is guaranteed by noticing that
$\phi^{+}\left( p;z\right) \propto p$ in the limit of $p \to 0$, as we have shown above.
Besides, the expression above also satisfies
\begin{eqnarray}
_{z}\langle z|g^{\left( +\right) }(k^{2}\to 0)|z^{\prime }\rangle_{z}
=
{}_{z}\langle z^{\prime }|g^{\left( +\right) }(k^{2} \to 0)|z\rangle _{z}^{\ast },
\end{eqnarray}
which indicates that the Green's operator converges to a Hermitian operator in
this low-energy limit.

Finally, we discuss the low-energy behavior of the scattering state
$\phi^{\left( -\right) }\left( k,z\right) $ with in-going boundary conditions.
The Lippmman-Schwinger equation for $\phi ^{\left( -\right) }\left( k,z\right) $
reads
\begin{eqnarray}
\phi ^{\left( -\right) }\left( k,z\right)  &=&\cos kz+
\int dz^{\prime} \Big[ g_{0}^{\left( -\right) }\left( k;z,z^{\prime }\right)
\nonumber \\
&& \hspace{1cm}
\times U\left( z^{\prime}\right) \phi ^{\left( -\right) }\left( k,z^{\prime }\right) \Big],
\label{OLPb}
\end{eqnarray}
where $g_{0}^{\left( -\right) }\left( k;z,z^{\prime }\right) $ is defined as
\begin{eqnarray}
g_{0}^{\left( -\right) }\left( k;z,z^{\prime }\right)  &=&
_{z}\langle z|\frac{1}{k^{2}-i0^{+}-{\hat T}_{z}}|z^{\prime }\rangle _{z}
\nonumber \\
&=&
i\frac{1}{2k}e^{ik\left\vert z-z^{\prime }\right\vert }. \label{g0fu}
\end{eqnarray}

It is easy to prove that, $\phi ^{\left( -\right) }\left( k,z\right) $ and
$\phi ^{\left( +\right) }\left( k,z\right) $ satisfy the same 1D
Schr\"{o}dinger equation with the same eigenvalue, and they also
share the same parity. Thus, $\phi ^{\left( -\right) }\left( k,z\right) $ is
proportional to $\phi ^{\left( +\right) }\left( k,z\right) $ and we have
\begin{eqnarray}
\phi ^{\left( -\right) }\left( k,z\right) =\xi \left( k\right) \phi ^{\left(
+\right) }\left( k,z\right) . \label{aa}
\end{eqnarray}
By expanding Eqs. (\ref{OLPa}) and (\ref{OLPb}) to the zeroth order of $k$,
and using Eqs. (\ref{g0}), (\ref{phizheng}), (\ref{g0fu}) and (\ref{aa}),
we obtain the following equations
\begin{eqnarray}
1-\frac{i}{2}\int dz^{\prime }\phi _{1i}\left( z^{\prime}\right)
U\left( z^{\prime }\right)  &=&0;  \label{ff} \\
1+\frac{i}{2}\xi \left( k\to 0\right) \int dz^{\prime}
\phi _{1i}\left( z^{\prime }\right) U\left( z^{\prime }\right)  &=&0,
\label{gg}
\end{eqnarray}
which leads to
\begin{eqnarray}
\xi \left( k\to 0\right) =-1
\end{eqnarray}
or
\begin{eqnarray}
\phi ^{\left( -\right) }\left( k,z\right) =
-\phi ^{\left( + \right) }\left(k,z\right)
\end{eqnarray}
in the limit of $k\to 0$.


\section{The Asymptotic Behavior of 1D Green's Operator}
\label{sec-appendix-C}

In this appendix we discuss the asymptotic behavior of the 1D Green's
operator. We focus on the case with even parity since it is the only relevant term
in our problem. In this case, the 1D Green's operator reads
\begin{eqnarray}
&&g_{\mathrm{even}}^{\left( +\right) }\left( k;z,z^{\prime }\right)
=
\nonumber \\
&& \hspace{5mm}
{}_{z}\langle z|\hat{P}_{\mathrm{even}}\frac{1}{k^{2}+i0^{+}
-{\hat T}_{z} - U\left( z\right) }
\hat{P}_{\mathrm{even}}|z^{\prime }\rangle _{z},
\label{r}
\end{eqnarray}
where $\hat{P}_{\mathrm{even}}$ is the projection operator to the space of
states with even parity of $z$.

We first investigate the asymptotic behavior of the free even Green's
function
\begin{eqnarray}
g_{\mathrm{even0}}^{\left( +\right) }\left( k;z,z^{\prime }\right) =
\hat{P}_{\mathrm{even}}g_{0}^{\left( +\right) }\left( k;z,z^{\prime }\right)
\hat{P}_{\mathrm{even}}  \label{s}
\end{eqnarray}
with $g_{0}^{\left( +\right) }\left( k;z,z^{\prime }\right) $ defined in (\ref{g0}).
In the case of $|z|>r_{\ast }$ and $|z'|<r_{\ast}$, we can show clearly that
\begin{eqnarray}
g_{\mathrm{even0}}^{\left( +\right) }\left( k;z,z^{\prime }\right) =
-i\frac{1}{2k}e^{ik|z|}\cos kz^{\prime }.  \label{t}
\end{eqnarray}

Now we consider the behavior of $g_{\mathrm{even}}^{\left( +\right) }\left(
k;z,z^{\prime }\right) $. We notice that $g_{\mathrm{even}}^{\left( +\right)
}\left( k;z,z^{\prime }\right) $ satisfies the Lippmman-Schwinger equation
\begin{eqnarray}
&&g_{\mathrm{even}}^{\left( +\right) }\left( k;z,z^{\prime }\right) =
g_{\mathrm{even0}}^{\left( +\right) }\left( k;z,z^{\prime }\right)   \notag
\label{u} \\
&&+\int dz^{\prime \prime }g_{\mathrm{even0}}^{\left( +\right) }\left(
k;z,z^{\prime \prime }\right) U\left( z^{\prime \prime }\right)
g_{\mathrm{even}}^{\left( +\right) }\left( k;z^{\prime \prime },z^{\prime }\right) .
\end{eqnarray}
Therefore, according to Eq. (\ref{t}), we know that in the region with
$|z|>r_{\ast}$ and $|z'|<r_{\ast}$ we have
\begin{eqnarray}
g_{\mathrm{even}}^{\left( +\right) }\left( k;z,z^{\prime }\right) =
- i\frac{1}{2k}e^{ik|z|}\chi \left( k,z^{\prime }\right) ^{\ast },
\end{eqnarray}
where
\begin{eqnarray}
\chi \left( k,z^{\prime }\right) ^{\ast } &=&\cos kz^{\prime }
+\int dz^{\prime \prime }
\Big[
\cos \left( kz^{\prime \prime }\right)
\nonumber \\
&& \hspace{1cm}
\times U\left( z^{\prime \prime }\right)
g_{\mathrm{even}}^{\left( +\right) }\left( k;z^{\prime \prime },z^{\prime }\right)
\Big].
\label{x}
\end{eqnarray}
Here, we have used the fact that $U\left( z\right) =0$ in the region $|z|>r_{\ast }$.
Notice that Eq. (\ref{x}) implies
\begin{eqnarray}
\chi \left( k,z^{\prime }\right)  &=&\cos kz^{\prime }
+\int dz \Big[
U\left(z^{\prime \prime }\right)
\nonumber \\
&& \hspace{1cm}
\times g_{\mathrm{even}}^{\left( -\right) }\left(k;z^{\prime \prime },z^{\prime }\right)
\cos kz^{\prime \prime }
\Big],
\label{y}
\end{eqnarray}
where the Green's function $g_{\mathrm{even}}^{\left( -\right) }
\left(k;z,z^{\prime }\right) $ is defined as
\begin{eqnarray}
&&g_{\mathrm{even}}^{\left( -\right) }\left( k;z,z^{\prime }\right)
=
\nonumber \\
&& \hspace{5mm}
{}_{z}\langle z|\hat{P}_{\mathrm{even}}\frac{1}{k^{2}-i0^{+}
-{\hat T}_{z}-U\left(z\right) }\hat{P}_{\mathrm{even}}|z^{\prime }\rangle _{z}.
\label{z}
\end{eqnarray}
It is apparent that the right-hand-side of Eq. (\ref{y}) is equivalent with
the Lippmman-Schwinger equation (\ref{OLPb}) for the scattering state $\phi^{-}(k,z)$ with
in-going boundary condition, leading to $\chi(k,z)=\phi^{-}(k,z)$.
Thus, in the region $|z|>r_{\ast }$ we have
\begin{eqnarray}
g_{\mathrm{even}}^{\left( +\right) }\left( k;z,z^{\prime }\right)
=
-i\frac{1}{2k}e^{ik\left\vert z\right\vert }
\phi ^{\left( -\right)}\left( k,z^{\prime }\right) ^{\ast }.
\label{asy}
\end{eqnarray}


\section{Background Scattering State}
\label{sec-appendix-D}

In this appendix we investigate the low energy behavior of the background
scattering state $|\psi_{\mathrm{bg}}^{\left( +\right) }\left( k\right) \rangle |o\rangle _{s}$,
the background Green's function $G_{\rm bg}^{(+)}$ and the
background scattering amplitude $f_{\rm even}^{\rm bg}$.
The state $|\psi _{\mathrm{bg}}^{\left( +\right) }\left(k\right) \rangle $
is given by the Lippmman-Schwinger equation (\ref{LPbg}).
For convenience we define the states
$|\psi _{\mathrm{bg}}^{\left( +\right)}\left( k\right) \rangle _{P,Q}$ as
\begin{eqnarray}
|\psi _{\mathrm{bg}}^{\left( +\right) }\left( k\right) \rangle _{P} &=&
{\hat P}_g|\psi _{\mathrm{bg}}^{\left( +\right) }\left( k\right) \rangle ,
\\
|\psi _{\mathrm{bg}}^{\left( +\right) }\left( k\right) \rangle _{Q}
&=&\left( 1-{\hat P}_g\right) |\psi _{\mathrm{bg}}^{\left( +\right) }\left(
k\right) \rangle,
\end{eqnarray}
where ${\hat P}_g = |00\rangle _{\bot }\langle 00|$ is the projection operator
to the transversally ground state.
Following a similar procedure as in Appendix \ref{sec-appendix-A},
we can write the Lippmman-Schwinger equation (\ref{LPbg})
as
\begin{eqnarray}
|\psi _{\mathrm{bg}}^{\left( +\right) }\left( k\right) \rangle _{P} &=&
|\phi_{P}^{\left( +\right) }\left( k\right) \rangle _{z}|00\rangle _{\bot}
\nonumber \\
&& \hspace{5mm}
+G_{P}^{\left( +\right) }\left( k\right)
{\hat X} |\psi _{\mathrm{bg}}^{\left(+\right) }\left( k\right) \rangle _{Q};
\nonumber\\ \label{LPbg1} \\
|\psi _{\mathrm{bg}}^{\left( +\right) }\left( k\right) \rangle _{Q}
&=&G_{Q}^{\left( +\right) }\left( k\right) {\hat X} |\psi _{\mathrm{bg}}^{\left(
+\right) }\left( k\right) \rangle _{P}  \label{LPbg2}
\end{eqnarray}
where
\begin{eqnarray}
G_{P}^{\left( +\right) }\left( k\right) &=&\frac{1}{k^{2}+i0^{+}-{\hat T}_{z}
+{\hat P}_g\left( {\hat H}_{\perp }+V_{\mathrm{bg}}\right) {\hat P}_g}; \\
G_{Q}^{\left( +\right) }\left( k\right) &=&\frac{1}{k^{2}-{\hat T}_{z}
+\left( 1-{\hat P}_g\right) \left( {\hat H}_{\perp }+V_{\mathrm{bg}}\right)
\left( 1-{\hat P}_g\right) }; \nonumber\\ \\
{\hat X} &=&{\hat P}_g\left({\hat H}_{\perp}+V_{\mathrm{bg}}\right) \left( 1-{\hat P}_g\right)
+h.c.,
\end{eqnarray}
and $|\phi _{P}^{\left( +\right) }\left( k\right) \rangle _{z}$ is the
1D scattering state with respect to the potential
$_{\bot}\langle 00|V_{\mathrm{bg}}\left( r\right) |00\rangle _{\bot }$
in the $z$ direction. We further rewrite Eq. (\ref{LPbg2}) as
\begin{eqnarray}
|\psi _{\mathrm{bg}}^{\left( +\right) }\left( k\right) \rangle
_{Q}=\sum_{i}|c_{i}\rangle d_{i}  \label{bgq}
\end{eqnarray}
with
\begin{eqnarray}
d_{i}=\frac{1}{k^{2}-\varepsilon _{i}}
\langle c_{i}|{\hat X}|\psi _{\mathrm{bg}}^{\left( +\right) }\left( k\right) \rangle _{P}  \label{di}.
\end{eqnarray}
Here, $|c_{i}\rangle $ is the eigenstate of the Hamiltonian
${\hat T}_{z}+\left( 1-{\hat P}_g\right) \left( {\hat H}_{\perp }+V_{\mathrm{bg}}\right) \left( 1-{\hat P}_g\right) $
with eigenenergy $\varepsilon _{i}$. Combining Eqs. (\ref{LPbg1}-\ref{LPbg2}) and
(\ref{bgq}-\ref{di}) together, we get a set of equations for the parameters $d_i$,
leading to the following expression for the state
$|\psi _{\mathrm{bg}}^{\left( +\right) }\left(k\right) \rangle _{P}$
\begin{eqnarray}
&&|\psi _{\mathrm{bg}}^{\left( +\right) }\left( k\right) \rangle _{P} =
|\phi_{P}^{\left( +\right) }\left( k\right) \rangle _{z}|00\rangle _{\bot }
\nonumber \\
&& \hspace{5mm}
+\sum_{i,j}G_{P}^{\left( +\right) }\left( k\right) {\hat X} |c_{i}\rangle D_{ij}^{-1}\left( k\right)
\langle c_{i}|{\hat X}|\phi _{P}^{\left( +\right)}\left( k\right) \rangle _{z}|00\rangle _{\bot },
\nonumber \\
&&
\label{psibg}
\end{eqnarray}
where $D_{ij}^{-1}\left( k\right) $ are the elements of the inverse of matrix $D\left( k\right) $
\begin{eqnarray}
D_{ij}\left( k\right) =\left( k^{2}-\varepsilon _{i}\right) \delta_{ij}
-\langle c_{i}|XG_{P}^{\left( +\right) }\left( k\right) {\hat X} |c_{j}\rangle.
\end{eqnarray}

Now we investigate the behavior of
$|\psi _{\mathrm{bg}}^{\left( +\right) }\left( k\right) \rangle _{P}$ in the limit of $k\to 0$.
As shown in Appendix \ref{sec-appendix-B}, in this low energy limit the 1D scattering state
$|\phi _{P}^{\left( +\right) }\left( k\right) \rangle _{z}$ is a linear function of $k$,
and the Green's function $G_{P}^{\left( +\right)}\left( k\right) $ becomes a
constant Hermitian operator. Substituting these properties into Eq. (\ref{psibg}),
we get
\begin{eqnarray}
|\psi _{\mathrm{bg}}^{\left( +\right) }\left( k\to 0\right) \rangle_{P}\propto k. \label{pbgp}
\end{eqnarray}
That is, the background scattering state
$|\psi _{\mathrm{bg}}^{\left(+\right) }\left( k\right) \rangle _{P}$
also linearly depends on $k$ in the low energy limit. Substituting (\ref{pbgp}) into (\ref{LPbg2}),
we get the low energy behavior of the  total background scattering state
\begin{eqnarray}
|\psi _{\mathrm{bg}}^{\left( +\right) }\left( k\to 0\right) \rangle \propto k.
\label{pbgtotal}
\end{eqnarray}
Using this result, and following the same analysis as in Appendix \ref{sec-appendix-B},
we can show that the background Green's operator
$G_{\mathrm{bg}}^{\left( +\right) }\left(k\right) $ defined in (\ref{ggbg})
converges to a Hermitian operator in the low-energy limit $k\to 0$.

In the above we obtained the low-energy behavior of the background scattering state
and the background Green's function.
Next we consider the low-energy behavior of the background scattering
amplitude. We also assume $V_{\mathrm{bg}}\left( r\right) =0$
when $r>r_{\ast }$. Then in the region $|z|\to \infty$,
the  wave function of $|\psi _{\mathrm{bg}}^{\left( +\right)}\left( k\right) \rangle _{Q}$
decays to zero while the wave function of
$|\psi _{\mathrm{bg}}^{\left( +\right)}\left( k\right) \rangle _{P}$ can be expanded as
\begin{eqnarray}
_{\bot }\langle 00|_{z}\langle z|\psi _{\mathrm{bg}}^{\left( +\right)}\left( k\right) \rangle _{P}
=\cos kz+f_{\mathrm{even}}^{\mathrm{bg}}\left( k\right) e^{ik\left\vert z\right\vert }
\end{eqnarray}
with the background scattering amplitude
\begin{eqnarray}
f_{\mathrm{even}}^{\mathrm{bg}}\left( k\right) &=&
f_{\mathrm{even}}^{P}\left( k\right) -i\frac{1}{2k}\sum_{ij}\text{ }
_{z}\langle \phi_{P}^{\left( -\right) }\left( k\right) |_{\bot }\langle 00|U|c_{i}\rangle
\nonumber \\
&&\times D_{ij}^{-1}\left( k\right) \langle c_{i}|U
|\phi _{P}^{\left(+\right) }\left( k\right) \rangle _{z}|00\rangle _{\bot }.
\label{feven}
\end{eqnarray}
Here, $f_{\mathrm{even}}^{P}\left( k\right) $ is the scattering amplitude for
the state $|\phi _{P}^{\left( +\right) }\left( k\right) \rangle _{z}$,
and $|\phi _{P}^{\left( -\right) }\left( k\right) \rangle _{z}$ is the
1D scattering state with respect to
$_{\bot }\langle 00|V_{\mathrm{bg}}\left( r\right) |00\rangle _{\bot }$
under the in-going boundary condition. To derive the equation above,
we have used the behavior of the 1D Green's function $G_{P}^{\left( +\right) }\left( k\right) $
in the region $z>r_{\ast}$ (see Appendix \ref{sec-appendix-C})
\begin{eqnarray}
&&
{}_{z}\langle z|_{\bot }\langle 00|G_{P}^{\left( +\right) }\left( k\right)
|00\rangle _{\bot }|z^{\prime }\rangle _{z}
=
\nonumber \\
&& \hspace{1.5cm}
-i\frac{1}{2k}e^{ik\left\vert
z\right\vert }{}_{z}\langle \phi _{P}^{\left( -\right) }\left( k\right)
|z^{\prime }\rangle _{z}.
\end{eqnarray}

As we have shown in Appendix \ref{sec-appendix-B}, 
the scattering amplitude $f_{\mathrm{even}}^{P}\left( k\right) $ and
scattering states $|\phi _{P}^{\left( \pm\right) }\left( k\right) \rangle _{z}$
have the following properties in the limit $k\to 0$
\begin{eqnarray}
f_{\mathrm{even}}^{P}\left( k\right) &=&-1+ika_{\mathrm{1D}}^{P}, \quad
a_{\mathrm{1D}}^{P}\in \mathrm{real} ;
\\
|\phi _{P}^{\left( +\right) }\left( k\right) \rangle _{z} &=&
-|\phi_{P}^{\left( -\right) }\left( k\right) \rangle _{z}.
\end{eqnarray}
Then it is obvious to see that Eq. (\ref{feven}) can be reformed as
\begin{eqnarray}
f_{\mathrm{even}}^{\mathrm{bg}}\left( k\right) =-1+ika_{\mathrm{1D}}^{\mathrm{bg}},
\end{eqnarray}
where the background scattering length
\begin{eqnarray}
a_{\mathrm{1D}}^{\mathrm{bg}} &=&a_{\mathrm{1D}}^{P}
+\lim_{k\to 0} \frac{1}{2k^2}\sum_{ij}
\text{ }_{z}\langle \phi _{P}^{\left( +\right)}\left( k\right) |_{\bot }\langle 00|U|c_{i}\rangle
\nonumber \\
&& \hspace{1cm}
\times D_{ij}^{-1}\left( k\right) \langle c_{i}|U
|\phi _{P}^{\left(+\right) }\left( k\right) \rangle _{z}|00\rangle _{\bot }
\end{eqnarray}
takes real value.

In the end of this appendix, we stress that the analysis in Appendix \ref{sec-appendix-C}
on the asymptotic behavior of the 1D Green's operator can be
straightforwardly generalized to the quasi-1D case and we have
\begin{eqnarray}
_{z}\langle z|G_{\mathrm{bg}}^{\left( +\right) }\left( k\right) =
-i\frac{e^{ik\left\vert z\right\vert }}{2k}|00\rangle _{\perp }
\langle \psi _{\mathrm{bg}}^{\left( -\right) }\left( k\right) |
\end{eqnarray}
in the region of $\left\vert z\right\vert >r_{\ast }$. Here, the scattering sate
$|\psi _{\mathrm{bg}}^{\left( -\right) }\left( k\right) \rangle $ with
in-going boundary condition is given by
\begin{eqnarray}
|\psi _{\mathrm{bg}}^{\left( -\right) }\left( k\right) \rangle =
|k\rangle_{z}|00\rangle _{\bot }+G_{\mathrm{bg}}^{\left( -\right) }\left( k\right)
V_{\mathrm{bg}}\left( r\right) |k\rangle _{z}|00\rangle _{\bot } \nonumber\\
\end{eqnarray}
with
\begin{eqnarray}
G_{\mathrm{bg}}^{\left( -\right) }\left( k\right) =
\frac{1}{k^{2}-i0^{+}-{\hat T}_{z}+{\hat H}_{\perp }+V_{\mathrm{bg}}}.
\end{eqnarray}
%



\begin{thebibliography}{99}

\bibitem{stoferle-04}
T. St{\"o}ferle, H. Moritz, C. Schori, M. K{\"o}hl, and T. Esslinger
Phys. Rev. Lett. {\bf 92}, 130403 (2004).

\bibitem{Kinoshita-04}
T. Kinoshita, T. Wenger, and D. S. Weiss, Science \textbf{305}, 1125 (2004);

\bibitem{Paredes-04}
B. Paredes \textit{et al.}, Nature \textbf{429}, 277 (2004);

\bibitem{Moritz-05}
H. Moritz, T. St\"{o}ferle, K. G\"{u}nter, M. K\"{o}hl, and T. Esslinger,
Phys. Rev. Lett. \textbf{94}, 210401 (2005);

\bibitem{Stock-05}
S. Stock, Z. Hadzibabic, B. Battelier, M. Cheneau, and J. Dalibard,
Phys. Rev. Lett. \textbf{95}, 190403 (2005);

\bibitem{Gunter-05}
K. G{\"u}nter, T. St{\"o}ferle, H. Moritz, M. K{\"o}hl, and T. Esslinger,
Phys. Rev. Lett. {\bf 95}, 230401 (2005).

\bibitem{Chin-06}
J.K. Chin \textit{et al.}, Nature \textbf{443}, 961 (2006);

\bibitem{Hadzibabic-06}
Z. Hadzibabic, P. Kr\"{u}ger, M. Cheneau, B. Battelier, and J. Dalibard,
Nature {\bf 441}, 1118 (2006);

\bibitem{Zimmermann-07} J. Fort\'{a}gh and C. Zimmermann, Rev. Mod. Phys. \textbf{79},
235 (2007), and references therein.

\bibitem{Syassen-08}
N. Syassen \textit{et. al.}, Science \textbf{320}, 1329 (2008).

\bibitem{Haller-09}
E. Haller \textit{et. al.}, Science \textbf{325}, 1224 (2009).

\bibitem{Lieb-63}
E.H. Lieb and W. Liniger, Phys. Rev. {\bf 130}, 1605 (1963).

\bibitem{Girardeau-63}
M. Girardeau, J. Math. Phys. (N.Y.) {\bf 1}, 516 (1906); Phys. Rev. {\bf 139}, B500 (1963).

\bibitem{Tonks-36}
L. Tonks, Phys. Rev. {\bf 50}, 955 (1936).

\bibitem{Girardeau-01}
M.D. Girardeau, E.M. Wright, and J.M. Triscari, Phys. Rev. A {\bf 63}, 033601 (2001).

\bibitem{Dunjko-01}
V. Dunjko, V. Lorent, and M. Olshanii, Phys. Rev. Lett. {\bf 86}, 5413 (2001).

\bibitem{Petrov-00a}
D.S. Petrov, G.V. Shlyapnikov, and J.T.M. Walraven, Phys. Rev. Lett. {\bf 85}, 3745 (2000).

\bibitem{Andersen-02}
J.O. Andersen, U. Al Khawaja, and H.T.C. Stoof, Phys. Rev. Lett. {\bf 88}, 070407 (2002);
U. Al Khawaja, J.O. Andersen, N.P. Proukakis, and H.T.C. Stoof, Phys. Rev. A {\bf 66}, 013615 (2002).

\bibitem{Kestner-06}
J.P. Kestner and L.-M. Duan,
Phys. Rev. A \textbf{74}, 053606 (2006).

\bibitem{Zhang-08}
W. Zhang, G.-D. Lin, and L.-M. Duan, Phys. Rev. A {\bf 77}, 063613 (2008);
{\it ibid}. {\bf 78}, 043617 (2008).

\bibitem{Olshanii-98}
M. Olshanii, Phys. Rev. Lett. {\bf 81}, 938 (1998).

\bibitem{Bergeman-03}
T. Bergeman, M. G. Moore, and M. Olshanii, Phys. Rev. Lett. {\bf 91}, 163201 (2003).

\bibitem{Petrov-00b}
D.S. Petrov, M. Holzmann, and G.V. Shlyapnikov, Phys. Rev. Lett. {\bf 84}, 2551 (2000).

\bibitem{Kim-05}
J.I. Kim, J. Schmiedmayer, and P. Schmelcher, Phys. Rev. A {\bf 72}, 042711 (2005).

\bibitem{Naidon-07}
P. Naidon, E. Tiesinga, W.F. Mitchell, and P. Julienne, New J. Phys. {\bf 9}, 19 (2007).

\bibitem{Saeidian-08}
S. Saeidian, V.S. Melezhik, and P. Schmelcher, Phys. Rev. A {\bf 77}, 042721 (2008).

\bibitem{Mora-04}
C. Mora, R. Egger, A.O. Gogolin, and A. Komnik, Phys. Rev. Lett. {\bf 93}, 170403 (2004);
C. Mora, R. Egger, and A.O. Gogolin, Phys. Rev. A {\bf 71}, 052705 (2005).

\bibitem{Mora-05}
C. Mora, A. Komnik, R. Egger, and A.O. Gogolin, Phys. Rev. Lett. {\bf 95}, 080403 (2005)

\bibitem{Granger-04}
B.E. Granger and D. Blume, Phys. Rev. Lett. {\bf 92}, 133202 (2004).

\bibitem{Haller-10}
E. Haller, M.J. Mark, R. Hart, J.G. Danzl, L. Reichsollner, V. Melezhik, P. Schmelcher, and H.-C. N{\"a}gerl,
Phys. Rev. Lett. {\bf 104}, 153203 (2010).


\bibitem{pseudopotential} K. Huang and C. N. Yang, Phys. Rev. {\bf 105}, 767 (1957).

\bibitem{Busch-98}
T. Busch, B.-G. Englert, K. Rzazewski, and M. Wilkens, Found. Phys. {\bf 28}, 549 (1998).

\bibitem{Idziaszek-06}
Z. Idziaszek and T. Calarco, Phys. Rev. A {\bf 74}, 022712 (2006).

\bibitem{Prudnikov-86}
A.P. Prudniko, Yu.A. Brychkov, and O.I. Marichev,
{\it Integrals and Series}, Vol. II (Gordon and Breach, New York 1986).

\bibitem{JMP1} B. R. Levy and J. B. Keller, J. Math. Phys. {\bf 4}, 54 (1963).

\bibitem{gao1} B. Gao, Phys. Rev. A {\bf 58}, 4222 (1998).

\bibitem{taylor}
J.R. Taylor, {\it  Scattering Theory}, Wiley, New York, 1972.

\bibitem{julian}
T. Kohler, K. Goral and P.S. Julienne,
Rev. Mod. Phys. {\bf 78}, 1311 (2006).

\bibitem{Peng-10}
S.-G Peng, S.S. Bohloul, X.-J. Liu, H. Hu, and P. Drummond,
arXiv:1005.2794v2.

\end{thebibliography}
\end{document}